\documentclass[usenatbib]{mnras}

\usepackage{newtxtext,newtxmath}
\usepackage{makebox}
\usepackage{graphicx,scalerel}
\usepackage{amsmath}
\usepackage{xcolor}

\usepackage[T1]{fontenc}

\DeclareRobustCommand{\VAN}[3]{#2}
\let\VANthebibliography\thebibliography
\def\thebibliography{\DeclareRobustCommand{\VAN}[3]{##3}\VANthebibliography}

\newcommand{\smalls}{\scriptscriptstyle}

\title[GCs in dSph galaxies]{The spatial distribution 
of globular clusters in dwarf spheroidal galaxies and the timing problem}

\author[S\'anchez-Salcedo \& Lora]{F.~J.~S\'anchez-Salcedo,$^{1}$
\thanks{E-mail: jsanchez@astro.unam.mx}
          and V. Lora$^{2}$
\\
$^{1}$Instituto de Astronom\'\i a, Universidad Nacional Aut\'onoma
de M\'exico,
P.O. Box 70-264, Ciudad Universitaria, 04510, Mexico City, Mexico \\
$^{2}$Instituto de Radioastronom\'{\i}a y Astrof\'{\i}sica, 
Universidad Nacional Aut\'onoma de M\'exico, Morelia,
Michoacan, Mexico}

\date{Accepted XXX. Received YYY; in original form ZZZ}

\pubyear{2021}

\begin{document}
\label{firstpage}
\pagerange{\pageref{firstpage}--\pageref{lastpage}}
\maketitle

\begin{abstract}
The dynamical friction timescale of massive globular clusters (GCs) in the inner 
regions of cuspy dark haloes in dwarf spheroidal (dSph) galaxies can be much
shorter than the Hubble time. This implies that a small fraction of the GCs is 
expected to be caught close to the centre of these galaxies.
We compare the radial distribution of GCs predicted in simple Monte Carlo models
with that of a sample of $38$ spectroscopically confirmed 
GCs plus $17$ GC candidates, associated mainly to low-luminosity 
dSph galaxies. If dark matter haloes follow an NFW profile,  
the observed number of off-center GCs at projected distances less than 
one half the galaxy effective radius 
is significantly higher than models predict. This timing problem can be viewed
as a fine-tuning of the starting GC distances. As a result of the short sinking 
timescale for GCs in the central regions,
the radial distribution of GCs is expected to evolve significantly during the next $1-2$ Gyr.
However, dark matter haloes with cores of size comparable to 
the galaxy effective radii can lead to a slow orbital in-spiral of GCs in the
central regions of these galaxies, providing a simple solution to the timing problem.
We also examine any indication of mass segregation 
in the summed distribution of our sample of GCs.

\end{abstract}


\section{Introduction}

Globular clusters (GCs) are believed to have formed within the high-pressure 
environments of the protogalaxies or during mergers of
galaxies \citep{elm97,kru12,kru14,lah19}.
In massive early-type galaxies, GCs are found to display a colour bimodality: red (metal-rich) and blue (metal-poor) GCs. 
The metal-rich GCs approximatelly
follow the galaxy stellar surface brightness profile, while the blue, metal-poor
GCs have typically a more extended spatial distribution.
This may indicate that metal-poor GCs were formed at early stages of galaxy 
formation, or they were accreted onto the host galaxy from tidally stripped dwarf 
satellite galaxies \citep[e.g.,][]{forrem18}. 

Some authors have pointed out that the radial distribution of GCs 
in elliptical galaxies, being less centrally concentrated than the halo stars,
may be a consequence of  the combination of different evolutionary processes, 
such as dynamical friction and tidal dissolution \citep[e.g.,][]{cap99,cap01}.
The fusion of GCs that sink into the central regions may lead to the formation of 
nuclear star clusters (NSCs) as those observed in many intermediate-mass galaxies, 
including the Milky Way \citep{cap08,ant12,gne14},
and in low-mass early-type galaxies \citep[e.g.,][]{tre75,lot01}.
For galaxies with absolute magnitudes in the range
$-19\leq M_{B}\leq -12$, the scaling relation between the luminosity of the
host galaxy and the luminosity of the NSC
is consistent with the GC merging scenario \citep{tur12,den14,car21}.

For a sample of dwarf galaxies containing $30$ dwarf irregular galaxies (dIrr), $2$
dE, $2$ dwarf spheroidal (dSph) galaxies and $2$ Magellanic spirals (Sm galaxies), 
\citet{geo09} find that the most luminous GCs tend to be more centrally located
\citep[see also][]{tud15}.
They argue that this luminosity segregation could be the result of the combined
effect of dynamical friction plus a preference for the formation of more 
massive GCs in the nuclear regions. 

The radial migration of GCs in dSph and dIrr galaxies could provide clues on the 
dark matter density profile in these galaxies and about the mechanisms for building
up NSCs \citep[e.g.,][]{lea20}. 
While the rotation curves of dwarf
galaxies favour dark matter haloes with a constant-density core, 
determinations of the value of the inner slope of the dark matter profile from
the stellar kinematics in pressure-supported dwarf galaxies is a delicate issue.
Many investigations have modelled the stellar kinematics of the classical Milky
Way dSph galaxies to infer their dark matter distribution. 
Most of the studies favour
a cuspy dark matter inner profile for Draco \citep{jar13,jar13b,rea18,rea19,hay20,mas20},
but a cored halo for Fornax \citep{wal11,amo12,pas18,rea19}.
In the case of Sculptor, the results have been conflicting; some favour a cuspy halo
\citep{ric14,mas18}, some favour a cored halo \citep[e.g.,][]{wal11}, and other conclude
that both are consistent with observations \citep[e.g.,][]{str18,gen18}. 
According to the models in \citet{hay20}, the classical 
dSph galaxies favour cuspy dark matter inner profiles, $\rho\propto r^{-\gamma}$, with
$\gamma$ between $0.4$ and $1.35$, albeit with large uncertainties
\citep[see also][]{rea19}.  A cored dark halo is consistent
with the available kinematic data in the case of Sextans, Sculptor and Fornax, within
$95$ percent confidence intervals.

Much effort has been devoted to place constraints on the dark matter density
distribution of Fornax dSph galaxy from the present-day distribution of its GCs
\citep{tre76,goe06,san06,ang09,
ino09,col12,arc16,bol19,leu20,mea20,bar21,sha21}.
A cored dark matter halo could explain the so-called timing problem, 
that is, why none of the five GCs in Fornax 
dSph galaxy have been sunk to the centre if the dynamical friction timescale, 
at least for two of them, is shorter than the age of the GCs. However, a cuspy
halo cannot be ruled out.
\citet{col12} find that if the GCs were formed within the tidal
radius of Fornax and the dark matter halo is cuspy, at least one GC 
is dragged to the centre in $1-2$ Gyr, except when selecting very particular initial
conditions of the orbits ($\sim 2$ per cent probability). They suggest that the Fornax
GCs could have been accreted via merger or tidal capture. 
On the other hand, \citet{arc16} find that the GCs could have been formed in situ
in a cuspy halo
if they are on circular orbits and their projected distances to the Fornax centre
are close to the three-dimensional (3D) distances. \citet{leu20} use a different approach;
they constrain the formation location of the GCs in Fornax, and argue that the present-day
positions of the GCs require a large dark matter core and a dwarf-dwarf merger.
\citet{mea20} compare the GC in-spiralling rates in cuspy and cored dark haloes
and conclude that the spatial distribution of GCs in Fornax cannot be used to
distinguish between cored and cuspy potentials.
In a recent work, \citet{sha21} suggest that the sinking timescale of the GCs in Fornax 
may be underestimated if the present-day mass of the dark matter halo derived from 
stellar kinematic analysis is used to compute it,
because Fornax dSph galaxy may have undergone significant mass stripping. 
Even more remarkably, the present-day distribution of GCs found
in the cosmological hydrodynamical simulation E-MOSAICS, which includes
the formation and evolution of GCs, is fully consistent with that of Fornax \citep{sha21}.

As a further step towards understanding the role of dynamical friction and its
effect on GCs orbiting dwarf galaxies, we investigate the spatial distribution of GCs 
in a sample of low-luminosity dwarf galaxies. 
Using a probabilistic approach, we explore if the main properties of the summed 
distribution of GCs in these galaxies can be accounted for, in a simple scenario where 
the orbits of the GCs decay towards the centres of galaxies due to dynamical friction 
with the dark matter particles in a cuspy halo or, on the contrary, it requires 
a finely tuned set of initial conditions. 

The paper is organized as follows. In Section \ref{sec:sample},
we describe our sample of GCs. In Section \ref{sec:correlations}, we search for 
any statistical correlation between relevant quantities. In particular, we investigate any
evidence of mass segregation of GCs. In Section \ref{sec:MC_model} we describe
the input parameters of simple Monte Carlo models, which are used as a tool
to interpret the present-day spatial distribution of GCs. In Section \ref{sec:modelA},
we compare the expected radial distributions of GCs with the observed one, assuming
that the dark matter haloes are cuspy.
In Section \ref{sec:past_future} we study the past and future spatial distributions 
of the GCs in these haloes. 
In Section \ref{sec:discussion} we show that the distribution of GCs 
is consistent with cored dark matter haloes.
There, we also discuss the limitations of our approach.
Finally, our main conclusions are summarized in Section \ref{sec:conclusions}.

\begin{figure*}
\includegraphics[width=182mm, height=80mm]{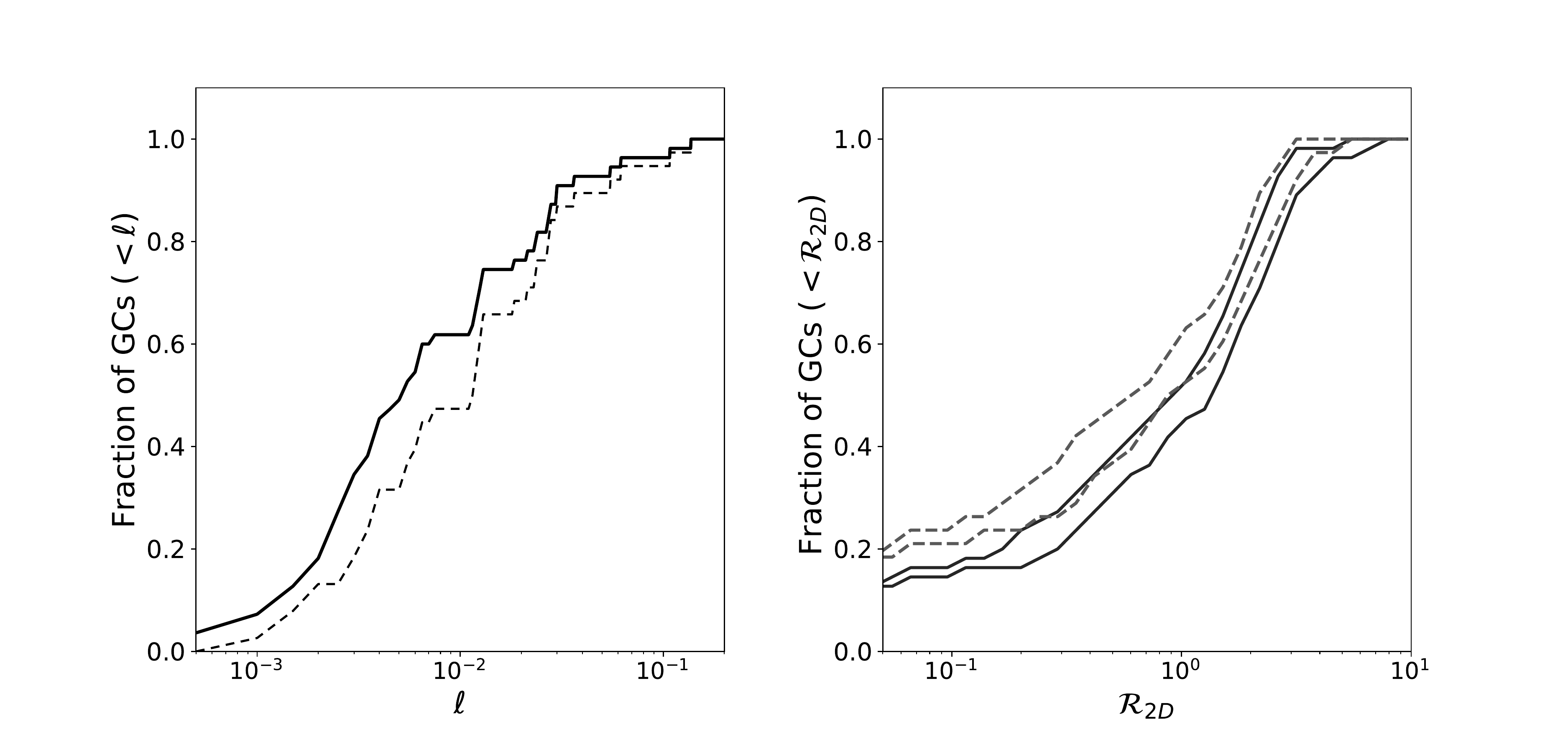} 
 \caption{
Cumulative distributions of the specific luminosity 
$\ell\equiv L_{\smalls V,\rm GC}/L_{\smalls V,\rm gal}$ (left panel) and 
the dimensionless projected radius $\mathcal{R}_{2\smalls \rm D}\equiv R_{\rm proj}/R_{e}$
(right panel) for the GCs in the extended sample (solid lines) and for confirmed
GCs (dashed lines). In the right panel, the curves delineate
the $2\sigma$ confidence bands.
}
 \label{fig:GCs_distribution_l_Rproj}
 \end{figure*}

\setcounter{table}{0}

\begin{table*}
\begin{center}
\caption{Fundamental parameters of galaxies}
\begin{tabular}{lccccrcc}\\ \hline \hline
 Name   &  Type &  $D$      & $M_{V}$& $R_{e}$& $\Theta$ & Number of GCs & Source  \\
	& & (Mpc) & & $(\arcsec)$ & & Total (confirmed)  &   \\
(1)        &    (2)& (3) & (4) & (5) & (6)  & (7) & (8) \\
\hline
Fornax dSph &  $-3$ &  0.147  & $-13.30$ & $996\pm 108$ & $2.3$ &5 (5) & M12, dB16  \\
Eridanus II &  $-3$ & 0.366 & $-7.1$ & $138.6\pm 7.2$ & $1.5$ &  1 (1) & Cr16, S21  \\
And I (KK 8) &  $-3$ & 0.745 & $-11.7$ & $186\pm 19$ & $3.7$ & 1 (0) & M12, Ca17 \\
And XXV &  $-3$ & 0.812 & $-9.7$ & $180\pm 15$ & $3.0$ & 1 (1) & Cu16 \\
DDO 216 (PegDIG) &  $10^{\star}$ & 0.920 & $-12.5$ & $156\pm 8.5$ 
&$1.2$&  1 (1)& M12, K14, Co17  \\
Sextans A (UGCA 205, DDO 75) & $10^{\ast}$ & 1.42 & $-14.2$ & $148\pm 14$ 
&$-0.6$& 1 (1) & M12, B14, B19 \\
KKs 3&  $-3$ & 2.12 & $-12.3$ & $50\pm 15$  &$-0.3$&  1 (1) & K15, S17  \\
KKs 55 & $-3$ & 3.94 & $-11.17$ & $50\pm 5$ & $3.1$&  1 (0)& K04, S08, G09 \\
KKs 58 & $-3^{\star}$ & 3.36 & $-11.93$ & $15.0\pm 0.5$ & $0.6$ & 1 (1) & F20 \\
Sc 22 (Scl-dE1)&  $-3$ & 4.3 & $-11.50$ & $35\pm 10$ & $0.9$ &  1 (1)& S08, D09 \\
IKN  &  $-3$ & 3.61 & $-11.5$ & $55\pm 16$ &  $2.7$ & 5 (5)& T15   \\
BK6N (KK 91) &  $-3$ &  3.85 & $-11.92$ &  $28\pm 12$& $1.1$ & 2 (0)& K04, S08, C09, S05  \\
KK 27 &  $-3$ & 3.98 & $-12.32$ & $25\pm 2$ &  $1.3$ & 1  (0)& K04, S08, S05  \\
KK 77 &  $-3$ & 3.48 & $-12.21$ & $34\pm 4$  & $2.0$ &  3 (0)& K04, S08, C09, S05  \\
KK 197 & $-3^{\star}$ & 3.87	& $-13.04$ & $38\pm 11$ & $3.0$  &  3 (3) &  S08, G09, F20 \\
KK 211 &  $-5$ & 3.58 & $-12.58$ & $26\pm 3$ & $1.5$ &  2 (2) & K04, S08, P08  \\
KK 221 & $-3$ & 3.98 & $-11.96$ & $32.3\pm 0.5$& $0.6$ & 6 (6)& K04, P08 \\
DDO 78 (KK 89) &  $-3$ & 3.72 & $-12.75$  & $38\pm 5$& $1.8$ &  2 (1)& K04, C09, L10, S03  \\
KDG 61 (KK 81) &  $-1$ & 3.60 & $-13.58$ & $43\pm 9$& $3.9$ &  1 (1) & K04, S08, C09, S05, M10  \\
KDG 63 (DDO 71, KK 83) &  $-3$ & 3.50  & $-12.82$ & $42\pm 16$& $1.8$ & 1 (1) & K04, S08, C09, S10  \\
F8D1 &  $-3$ & 3.77  & $-13.14$ & $67\pm 7$ & $2.0$ &  1 (0) & Ca98, C09, K00  \\
ESO 269-66 (KK 190) &  $-1$ & 3.82 & $-13.89$ & $40\pm 12$ & $1.7$ & 4 (1)& S08, G09, S17  \\
ESO 294-010 &  $-3^{\star}$ & 1.92 & $-11.40$ & $25\pm 2$ &  $1.0$ &1 (0) & K04, S08, S05 \\
ESO 384-016 &  $10^{\star}$ & 4.53 & $-13.72$ & $21\pm 2$ &  $0.3$ & 2 (0) & G09, dS10, G10 \\
ESO 540-030 (KDG 2, KK 9) &  $-1$ &  3.40 & $-11.84$ & $30\pm 4$ & $0.4$ &  1  (0) & K04,
S08, S05  \\
KK 84 &  $-3$ & 9.7 & $-14.40$ & $19\pm 2$ & $4.5$ &  6 (6) & K04, S08, P08 \\
\hline \hline
\end{tabular}
\end{center}
{\bf Notes}: 
Columns contain the following data:
{\bf (1)} galaxy name,
{\bf (2)} de Vaucouleurs $T$ morphological type according to \citet{kar04}
(asterisks indicate dSph/dIrr transition-type galaxies),
{\bf (3)} distance,
{\bf (4)} absolute $V$ magnitude,
{\bf (5)} effective radius,
{\bf (6)} tidal index,
{\bf (7)} total number of potential GCs and spectroscopically confirmed GCs,
{\bf (8)} 
References -- (B09) Beasley et al. (2019); (B14) Bellazzini et al. (2014); 
(Ca98) Caldwell et al. (1998); (Ca17) Caldwell et al. (2017); (C09) Chiboucas et al. (2009); 
(Co17) Cole et al. (2017); (Cr16) Crnojevi\'c et al. (2016); (Cu16) Cusano et al. (2016); 
(dB16) de Boer \& Fraser (2016); (dS10) de Swardt et al. (2010); (D09) Da Costa et al. (2009); 
(F20) Fahrion et al. (2020); (G09) Georgiev et al. (2009); (G10) Georgiev et al. (2010);
(K00) Karachentsev et al. (2000); (K04) Karachentsev et al. (2004);
(K15) Karachentsev et al. (2015); (K14) Kirby et al. (2014); (L10) Lianou et al. (2010);
(M10) Makarova et al. (2010);
(M12) McConnachie (2012); (P08) Puzia \& Sharina (2008); 
(S03) Sharina et al. (2003); (S05) Sharina et al. (2005); (S08) Sharina et al. (2008);
(S10) Sharina et al. (2010); (S17) Sharina et al. (2017); (S21) Simon et al. (2021);
(T15) Tudorica et al. (2015)\\
\label{table:dSph_properties}
\end{table*}

\begin{figure}
\includegraphics[width=89mm, height=76mm]{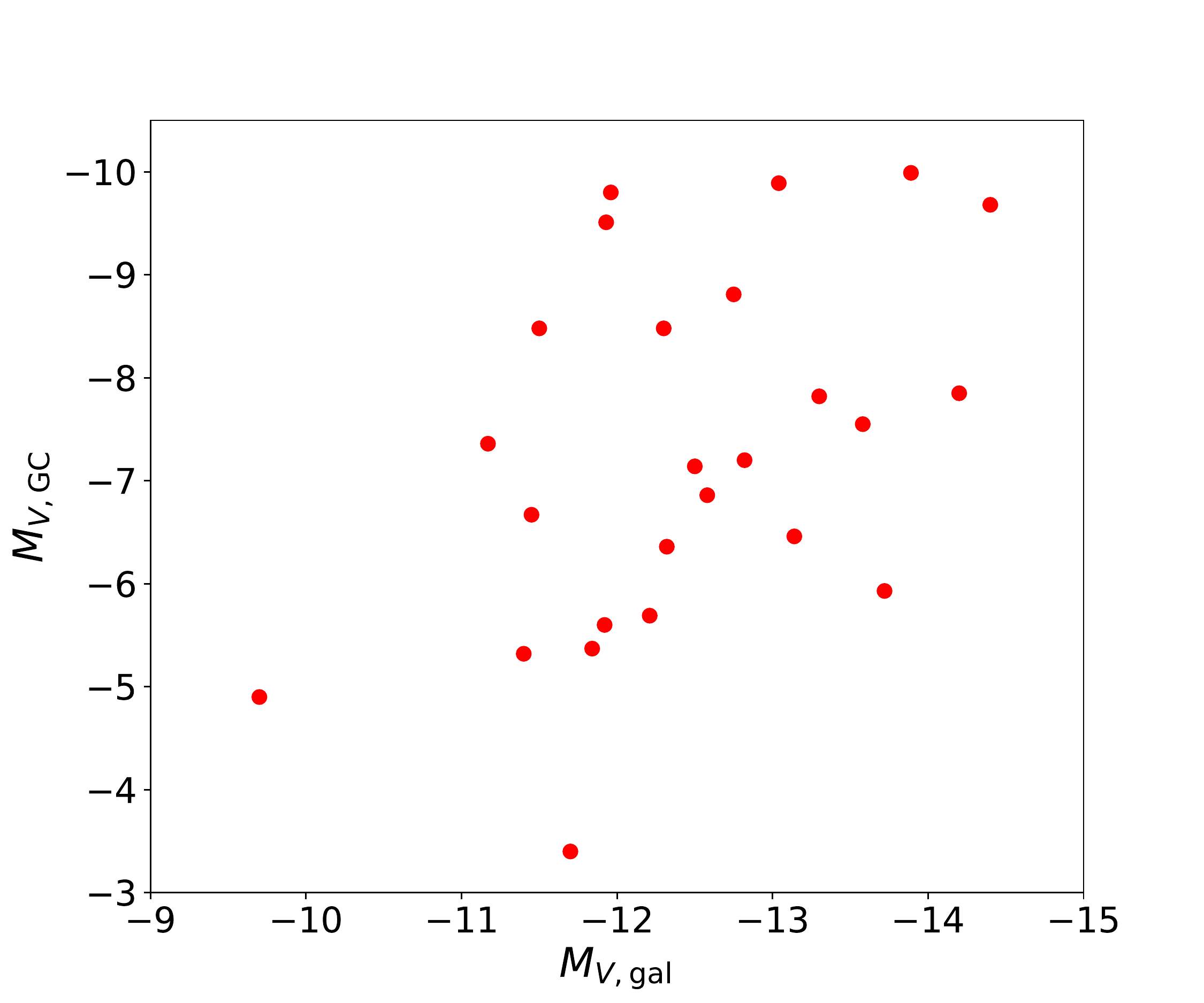}
 \caption{Relation between the absolute $V$ magnitude of the brightest GC
 and $M_{\smalls V,\rm gal}$ in the
extended sample. }
 \label{fig:MV_vs_MV}
 \end{figure}

\section{Sample of GCs in dwarf galaxies}
\label{sec:sample}

We gather a sample of $38$ spectroscopically confirmed GCs plus $17$ candidates,
around $26$ galaxies classified as dSph 
galaxies or transition-type galaxies (with properties intermediate between dIrr and
dSph galaxies), using available data from the literature.  
Some relevant properties of the selected galaxies, such as their $V$ band absolute
magnitudes $M_{\smalls V,\rm gal}$, and their effective radii $R_{e}$
(i.e. the projected half-light radii), are given in Table \ref{table:dSph_properties}. 
We see that $M_{\smalls V,\rm gal}$ ranges between $-7.1$ and $-14.4$. 
We do not have the formal errors on $M_{\smalls V,\rm gal}$ for all galaxies, but they are known with an accuracy less than $0.3$ mag for Local Group dwarfs,
and up to $0.5$ mag for low-surface brightness dSph galaxies outside the Local Group.
We also provide the so-called tidal index $\Theta$, taken from \citet{kar04}, 
which is a measure of the level of isolation of the galaxies (see Section \ref{sec:discussion}).
Note that the Sagittarius dwarf galaxy has not been included because it is 
currently undergoing tidal disruption.

\begin{figure}
\includegraphics[width=89mm, height=76mm]{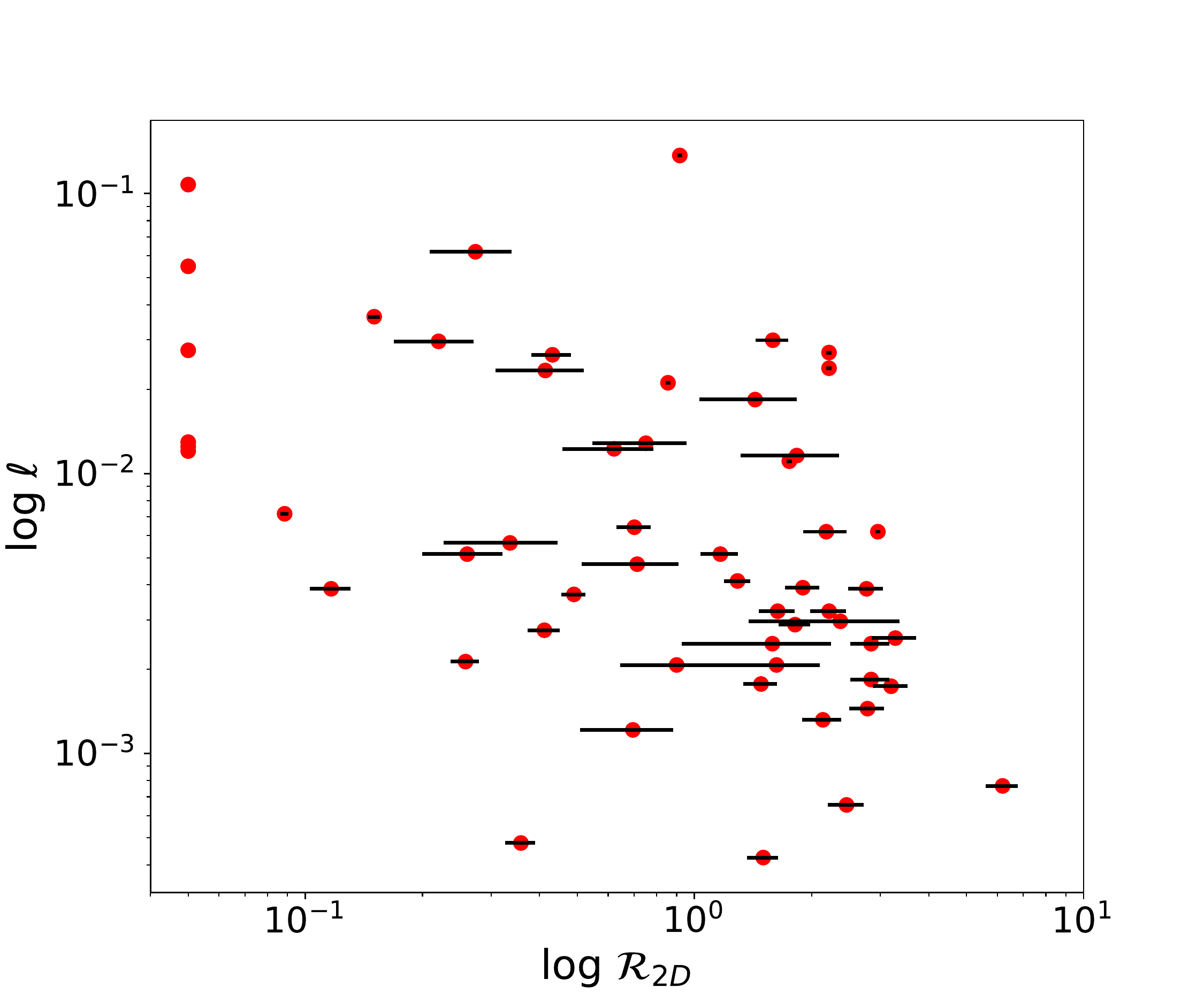}
 \caption{Relation between $\ell$ and $\mathcal{R}_{2\smalls \rm D}$ for GCs in the
extended sample. For plotting purposes, in order to make the graph
in logarithmic scale, we assigned a value of $\mathcal{R}_{2\smalls \rm D}=0.05$ to 
the central GCs. Typical fractional errors on $\ell$ are $\sim 30\%$.}
 \label{fig:l_vs_R2D}
 \end{figure}

Table \ref{table:GC_properties} compiles the $V$ band magnitude, 
$M_{\smalls V,{\rm GC}}$, of each GC in our sample, as well as their projected 
distances from the host centre (denoted by $R_{\rm proj}$) and their estimated ages.
In general, the typical uncertainty on $M_{\smalls V,{\rm GC}}$ 
is less than $0.1$ mag.
As already said, we have $17$ GC candidates for which
spectroscopic observations are required to rule out contamination by
stellar interlopers projected onto the dSph galaxy, or by 
background galaxies \citep[e.g.,][]{dac09}. Still, there is little question that some
of them (e.g., the one in KKs 55 or those in ESO 384-016) are real GCs 
\citep{geo10,for18}. For shortness, we will refer to the total sample of 
confirmed GCs plus candidates as the ``extended'' sample. 

Hereafter, we will refer to {\it central} or NSCs as those sit at the 
centre of the host galaxy, i.e.  $R_{\rm proj}=0$. In the sample of GCs under 
consideration, we have $6$ central GCs. And XXV Gep I is also considered 
as a central GC, although it is still unclear whether it is a central GC 
or the nucleus of And XXV \citep{cus16}. 

\begin{table}
\begin{center}
\caption{Fundamental parameters of GCs in the extended sample}
\begin{tabular}{lrcc}\\ \hline \hline
 Name/ID  &  $M_{V}$ &  $R_{\rm proj}$       &  Age  \\
	& & $(\arcsec)$ &    (Gyr)    \\
(1)        &    (2)& (3) & (4)      \\
\hline
Fornax GC 1$^{\star}$ &  $-5.34$ &  2400  &  $12.1\pm 0.8$     \\
Fornax GC 2$^{\star}$& $-7.07$  &  1580  &   $12.2\pm 1.0$    \\
Fornax GC 3$^{\star}$&  $-7.82$ &  648  &   $12.3\pm 1.4$    \\
Fornax GC 4$^{\star}$&  $-6.90$ &  360  &    $10.2\pm 1.2$    \\
Fornax GC 5$^{\star}$&  $-7.07$ &  2160  &  $11.5\pm 1.5$     \\
Eridanus II GC$^{\star}$ &  $-3.0$ & 13.9 &    $12.5\pm 1.5$\\
And I GC &  $-3.4$ & 57 &   --   \\
And XXV Gep I $^{\star}$&  $-4.9$ & $\leq 11$ &     --   \\
DDO 216-A1$^{\star}$ &  $-7.14$ & 6 &    $12.3\pm 0.8$  \\
Sextans A-GC1$^{\star}$ & $-7.85$ & 261 &   $8.6\pm 2.7$  \\
KKs 3 GC$^{\star}$&  $-8.48$ & 8.5 &    12.6  \\
KKs 55-01 & $-7.36$ & 77 &    -- \\
KKs 58-NSC$^{\star}$ & $-9.51$ & 0.0 &   $6.9\pm 1.0$  \\
Sc 22 (Scl-dE1) GC1$^{\star}$ &  $-6.67$ & 20 &    --\\
IKN-01$^{\star}$  &  $-6.66$ & 98 &   14.77  \\
IKN-02$^{\star}$ &  $-7.16$ & 76 &     --\\
IKN-03$^{\star}$  &  $-6.77$ & 38.5 &   13.21   \\
IKN-04$^{\star}$  &  $-7.42$ & 20 &   14.19   \\
IKN-05$^{\star}$  &  $-8.48$ & 12.3 &   13.80   \\
BK6N 2-524 &  $-5.40$ &  43 &   -- \\
BK6N 4-789 &  $-5.60$ &  65 &   --  \\
KK 27 4-721& $-6.36$ & 31 &   --   \\
KK 77 4-939&  $-5.01$ & 71 &    --\\
KK 77 4-1162&  $-5.37$ & 95 &    --\\
KK 77 4-1165&  $-5.69$ & 95 &    -- \\
KK 197-01$^{\star}$ & $-5.75$ & 24.5	&    --  \\
KK 197-02$^{\star}$ & $-9.89$ & 0	&   $6.5\pm 1$ \\
KK 197-03$^{\star}$& $-7.32$ & 8	&    $7\pm 1$  \\
KK 211 3-917$^{\star}$&  $-6.86$ & 29 &    $6\pm 2$ \\
KK 211 3-149$^{\star}$&  $-7.82$ & 0 &    $6\pm 2$ \\
KK 221 2-608$^{\star}$& $-8.04$ & 70 &   --   \\
KK 221 2-883$^{\star}$& $-7.07$ & 55 &   --   \\
KK 221 2-966$^{\star}$& $-9.80$ & 28 &   $10\pm 2$   \\
KK 221 2-1090$^{\star}$& $-7.77$ & 26 &  --   \\
KK 221 24n$^{\star}$& $\sim -7.9$ & 70 &  $9\pm 2$  \\
KK 221 27n$^{\star}$& $\sim -6.4$ & 94 &   --  \\
DDO 78 1-167 &  $-7.23$ & 81 &   --     \\
DDO 78 3-1082$^{\star}$ &  $-8.81$ & 14.5 &  $10.5\pm 1.5$    \\
KDG 61 3-1325$^{\star}$ & $-7.55$ & 2.86 &   $16\pm 2$    \\
KDG 63 3-1168$^{\star}$ &  $-7.2$ & 12  &    $6\pm 2$  \\
F8D1 GC&  $-6.46$ & 14  &    --   \\
ESO 269-66-01 &  $-8.08$ & 26.5 &    --   \\
ESO 269-66-03$^{\star}$ &  $-9.99$ & 0 &   $12.6$   \\
ESO 269-66-04 &  $-7.18$ & 63 &    --   \\
ESO 269-66-05 &  $-7.18$ & 34 &    --   \\
ESO 294-010 3-1104 &  $-5.32$ & 11 &   --   \\
ESO 384-016-01 &  $-5.93$ & 129 &   --    \\
ESO 384-016-02 &  $-5.29$ & 30.5 &   --   \\
ESO 540-030 GC &  $-5.37$ &  97 &     --   \\
KK 84 2-785$^{\star}$&  $-7.30$ & 52.5 &    --   \\
KK 84 3-705$^{\star}$&  $-8.38$ & 35.5 &    $9\pm 1$  \\
KK 84 3-830$^{\star}$&  $-9.68$ & 0 &    $10\pm 4$   \\
KK 84 3-917$^{\star}$&  $-7.52$ & 27.5 &  --  \\
KK 84 4-666$^{\star}$&  $-8.37$ & 52.2 &   $8\pm 3$   \\
KK 84 12n$^{\star}$&  $\sim -7.5$ & 60.5 &    --  \\
	
\hline \hline
\end{tabular}
\end{center}
{\bf Notes}: Columns contain the following data:
{\bf (1)} name or ID,
{\bf (2)} absolute $V$ magnitude,
{\bf (3)} projected distance to the centre of the host galaxy in arcsec,
{\bf (4)} age.
Asterisks after the name indicate those GCs that are spectroscopically confirmed.
This data was collected from the references given in Table \ref{table:dSph_properties}.
\\
\label{table:GC_properties}
\end{table}

We define the {\it specific} luminosity of a certain GC as the ratio between
its luminosity $L_{\smalls V, \rm GC}$ and the luminosity of the host galaxy
$L_{\smalls V, \rm gal}$, that is 
$\ell\equiv L_{\smalls V,\rm GC}/L_{\smalls V,\rm gal}$.
Figure \ref{fig:GCs_distribution_l_Rproj} shows the 
distribution of the specific luminosity, and the 
distribution of ${\mathcal{R}}_{2\smalls \rm D}\equiv R_{\rm proj}/R_{e}$ in both the
extended sample and for only confirmed GCs. 
The distributions of $\ell$ and ${\mathcal{R}}_{2\smalls \rm D}$ of the GCs in the 
extended sample are slightly different to those derived using the sample of 
confirmed GCs because all the central GCs have been confirmed
spectroscopically. Aside from that, the distributions are very similar.

Throughout this paper, we only include the uncertainties
associated with the determination of the effective radius $R_{e}$ of the galaxies.
Uncertainties in the distance to the galaxies and those associated
with photometry measurements (e.g., in $M_{\smalls V,\rm gal}$ and 
$M_{\smalls V,\rm GC}$) are not taken into account.

\section{Searching for correlations}
\label{sec:correlations}

Correlations between different physical quantities are very useful to detect possible
observational bias and to constrain suitable models. 
To evaluate possible correlations, we perform 
Spearman’s rank correlation tests. We consider that a correlation is strong 
when the Spearman’s coefficient $\rho_{s}$ is $\geq 0.6$, moderate when 
$0.4 \leq \rho_{s} < 0.6$, and weak when $\rho_{s} < 0.4$.
In addition to $\rho_{s}$, it is customary to give the $p$-value.
A given correlation is considered significant if the probability 
($p$-value) of getting a specific correlation coefficient by chance is lower than 
$5\%$ (i.e., $p < 0.05$). However, when the sample is small, the information in
the $p$-values is not complete. In such cases, it is more informative to estimate
the $z$-score or the confidence level of the Spearman coefficient \citep{cur15}.
Along this paper, we provide the $z$-scores but also the $p$-values for those
readers that are more familiar with this indicator.

We first examine any possible correlation between the magnitude of the host 
galaxy $M_{\smalls V,{\rm gal}}$ and the magnitude $M_{\smalls V,{\rm GC}}$ of 
its most luminous GC (see Fig. \ref{fig:MV_vs_MV}). For GCs in the extended sample,
the Spearman’s rank correlation coefficient is $\rho_{s}=0.52$
at a significance of $(3.0\pm 1.1)\sigma$ ($p$-value$<0.007$). This implies that there is
a positive, albeit moderate, correlation between $M_{\smalls V,{\rm gal}}$ and 
the magnitude of the most luminous GCs.
Assuming a constant 
stellar mass-to-light ratio for the galaxies and another constant value for the GCs, 
this relation is equivalent to the relation between the maximum GC mass and
the host galaxy stellar mass.
\citet{lea20} have looked at this relation for a sample of Local Group galaxies with 
stellar masses up to $10^{11}M_{\odot}$. In the range of low stellar-mass galaxies 
(masses below $10^{8}M_{\odot}$), we find a similar scaling relation.

We have also performed the Spearman analysis for the correlation between
$L_{\smalls V,{\rm gal}}$ and the total luminosity of the GC population
in each galaxy, and found $\rho_{s}=0.50$ with a significance $z$-score of 
$(2.8\pm 1.2)\sigma$, using the extended sample. The $p$-value is $<0.01$. 
The correlation is moderate because of the short range in galaxy luminosities.
A more clear positive trend between galaxy stellar mass and GC system mass 
is found when a wider range in galaxy luminosities is considered \citep[e.g.,][]{for18,lea20}.

We have also searched for a correlation between the luminosity of the GCs and
its distance to the host centre. 
As already said in the Introduction, there is some indications of mass segregation
of GCs in dwarf galaxies. Mass segregation of GCs may be primordial if
the most massive GCs were formed 
in the central regions, where converging flows of gas lead to reach high pressure.
Dynamical friction of GCs against dark matter and field stars
may also induce mass segregation since the
dynamical friction time scales as the inverse of the GC mass.

The correlation between $L_{\smalls V, \rm GC}$ and 
the dimensionless projected radius 
$\mathcal{R}_{2\smalls \rm D}$ is $-0.28\pm 0.15$ for the GCs in the extended
sample. If central GCs are excluded, the correlation is $-0.15\pm 0.15$.
The correlation increases if the specific luminosity $\ell$ is considered instead
of $L_{\smalls V, \rm GC}$. For the GCs in the extended sample,
we find a relatively moderate anticorrelation with a coefficient of 
$\rho_{s}=-0.45\pm 0.12$, at a significance level of $(3.5\pm 1.1)\sigma$.
The $p$-value is $6\times 10^{-4}$.  A plot of $\ell$ versus 
$\mathcal{R}_{2\smalls \rm D}$ is shown in Figure \ref{fig:l_vs_R2D}.
If we exclude the six NSCs,  
the correlation coefficient decreases to $-0.36\pm 0.14$.
On the other hand, for the sample of confirmed GCs, we find $\rho_{s}=-0.46\pm 0.13$
between $\ell$ and $\mathcal{R}_{2\smalls \rm D}$, at 
a significance level of $(3\pm 1)\sigma$.

Since $R_{e}$ in our sample of galaxies only varies by a factor of $\sim 5$, being
ESO 294-010 the galaxy with the lower $R_{e}$ ($233$ pc) and F8D1 with the largest 
$R_{e}$ ($1230$ pc), we have examined the correlation between $\ell$ and $R_{\rm proj}$
in parsecs. The difference between using $R_{\rm proj}$ or ${\mathcal{R}}_{2\smalls \rm D}$ 
is that the first one does not require knowledge of $R_{e}$, although is more sensitive to 
uncertainties in the adopted distances to the galaxies (which are not included here). 
We find that the strength of the correlation between $\ell$ and $R_{\rm proj}$ 
in the extended sample is $\rho_{s}=-0.45\pm 0.11$, with a $z$-score of 
$(3.5\pm 1.0)\sigma$. Remarkably, this correlation has the same strength as
the $\ell$-${\mathcal{R}}_{2\smalls \rm D}$ correlation.

An inspection of the GCs in our sample indicates that 
the GC named KK 221-24n is very close to a bright foreground star (see the image in
Fig. 2 of Puzia \& Sharina 2008), making it difficult to obtain reliable photometry. 
If this GC is excluded of our extended sample, the strength of the anticorrelation 
between $\ell$ and $\mathcal{R}_{2\smalls D}$ increases to $\rho_{s}=-0.5$ 
at a significance of $(3.7\pm 1.1)\sigma$.

\section{Monte Carlo model}
\label{sec:MC_model}

GCs gradually sink towards the host centre due to the dynamical friction against 
the dark matter particles. In the central regions, the strength of dynamical friction 
depends on the underlying dark matter profile. 
For instance, in cored dark matter haloes,
the rate of their inspiralling slows down when GCs reach the core radius 
\citep{mea20}, or may potentially even stall \citep{goe06,ino09,col12,pet16,leu20}. 
Therefore, the survival of GCs against orbital decay, their current positions, and
the potential coalescence of multiple GCs to form NSCs depend on the 
dark matter profile. Indeed, some authors have put constraints on the dark matter 
density profile and on the initial galactocentric distance of the GCs 
in Fornax \citep[e.g.,][]{ang09,arc16,mea20,sha21}, DDO 216 [Pegasus dIrr galaxy] \citep{col17,lea20} and Eridanus II \citep[e.g.,][]{amo17,con18}. 

As said in the Introduction, there is no consensus about whether
the timing problem of the orbital decay of Fornax GCs implies that its dark matter 
halo has a constant-density core. According to \citet{mea20}, 
the Fornax core radius required to solve the timing problem should be 
implausibly large. They suggest that the simplest explanation is that Fornax
GCs were formed at radii of $\sim 2$ kpc 
(outside Fornax effective radius), and are now on their way to sinking to the centre
\citep[see also][]{ang09,sha21}.  
Did all massive GCs in dSph galaxies form in their outskirts?

In order to investigate further whether cuspy dark haloes are 
consistent with the current radial distribution of the observed GCs,
we will assume that the dark matter haloes around dwarf galaxies are spherical, 
following an NFW profile, as predicted in cosmological $N$-body simulations:
\begin{equation}
\rho_{\rm dm}(r) = \frac{\rho_{0} r_{s}}{r\left(1+r/r_{s}\right)^{2}},
\end{equation}
where $\rho_{0}$ and $r_{s}$ are scale parameters that vary from galaxy to galaxy.
Nevertheless, we refer the reader to Section \ref{sec:discussion} for a discussion about the
GC in-spiralling in cored dark matter haloes.

\begin{figure*}
\includegraphics[width=177mm, height=55mm]{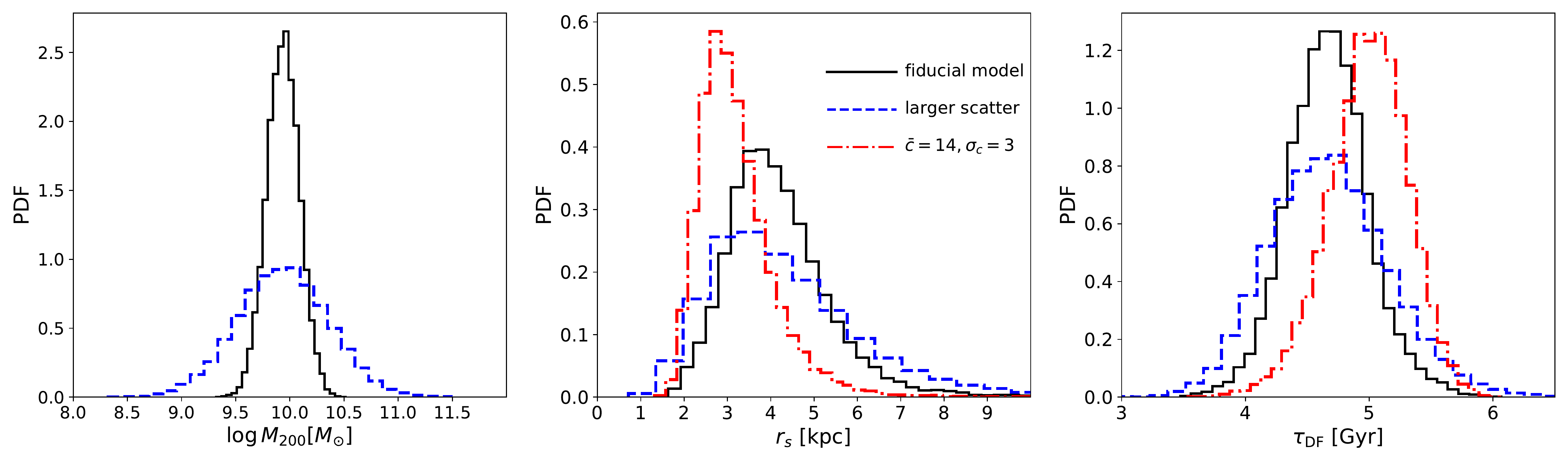}
 \caption{Probability distribution functions (PDFs) of $M_{200}$ (left panel), 
$r_{s}$ (middle panel) and
$\tau_{\rm \smalls DF}$ (right panel) for a galaxy with $M_{V}=-13.3$, which
corresponds to the Fornax $V$ band magnitude. 
Shown are histograms for our fiducial model (black solid lines), for a model
where the scatter in the $M_{\star}$-$M_{200}$ relation is three times 
the nominal scatter in \citet{rea17} (blue dashed histograms),
and for a model where the distribution of $c$ is
normal around a mean value $\bar{c}=14$ and standard deviation $\sigma_{c}=3$
(red dash-dotted histograms). $\tau_{\smalls \rm DF}$ is evaluated for a GC with
$M_{\rm \smalls GC}=2\times 10^{5}M_{\odot}$ at an initial 3D distance of $1$ kpc.
}
\label{fig:Fornax}
\end{figure*}

\citet{arc15} provide a useful interpolation formula for the dynamical friction timescale 
of massive point particles orbiting in spherical cuspy (or cored) density profiles
with isotropic velocity distribution functions.
The formula was calibrated against $N$-body models \citep{arc14}. 
In their fitting process, the minimum impact parameter in the 
Coulomb logarithm varies along the orbit to properly account for the magnitude
of the drag in the central parts where the density diverges and the local approximation
overestimates it.
In the particular case of an NFW profile, a GC with mass 
$M_{\smalls \rm GC}$ in an eccentric orbit with an initial apocentre
$r_{\rm \smalls apo,i}$ reaches the centre after a time
\begin{equation}
\tau_{\smalls \rm DF}[{\rm Myr}]= 0.3 g(\tilde{e}) \left(\frac{r_{s}}{1\,{\rm kpc}}\right)^{3/2}
\left(\frac{M_{\rm gal}}{10^{11}M_{\odot}}\right)^{-1/2}
\left(\frac{M_{\smalls \rm GC}}{M_{\rm gal}}\right)^{-0.67} \left(\frac{r_{\rm \smalls apo,i}}{r_{s}}\right)^{1.76},
\label{eq:tau_DF_Arca}
\end{equation}
where $M_{\rm gal}$ is the mass of the host galaxy, 
$\tilde{e}\equiv 1-(r_{\rm \smalls per,i}/r_{\rm \smalls apo,i})$ and
$g(\tilde{e})\simeq 5-4\tilde{e}$. Here $r_{\rm \smalls per,i}$ is the initial
pericentre. We will take the virial mass $M_{200}$ as a proxy for the mass
of the host galaxy. Note that this expresion for the sinking time 
does not take into account mass loss of the GCs.

Assuming that the eccentricity is preserved during orbital decay, the apocentre 
of the GC  decays to a value
\begin{eqnarray}
r_{\rm \smalls apo,f}=
\left(1-\frac{\tau}{\tau_{\rm \smalls DF}}\right)^{0.57} 
r_{\rm \smalls apo,i}
\label{eq:final_radius_df}
\end{eqnarray}
after a time $\tau$ \citep[e.g.,][]{arc16}.
Therefore, if we know the mass, the initial orbital parameters and the time 
at which the GCs started their orbital decay, as well as the scale parameters of the 
host dark matter haloes, we can compute the present-day orbital parameters of the GCs.
If two or more GCs reach the centre in a certain galaxy, 
then we assume that they merge together and sum their luminosities. 
Conversely, Equation (\ref{eq:final_radius_df}) allows to derive the 
past orbital parameters of the GCs if we knew their current values plus also the mass and
scale radius of the dark matter haloes.

The validity of Equation (\ref{eq:final_radius_df}) breaks down close to the centre 
of the host galaxy, typically at the galactocentric distance where the galaxy enclosed 
mass within the GC orbit is comparable to the GC mass
\citep[e.g.][]{gua08,goe10,arc14}. 
Indeed, this process may hinder the in-spiral of GCs in galaxies with shallow baryonic and total mass density profiles.
In order to isolate the different
effects and to properly intepret the results, we will omit this effect in Sections
\ref{sec:modelA} and \ref{sec:past_future}. The reader is referred to Section \ref{sec:discussion} for a discussion about the impact of this effect.

In order to estimate the parameters of the dark haloes of the sample galaxies,
we first calculate the stellar mass of the galaxies, $M_{\star}$, 
from their luminosity assuming a uniform distribution of the stellar 
mass-to-light ratio $(M_{\star}/L)_{\smalls \rm gal}$ between 
$1$ and $2$ \citep{mar05,mcc12}. Once $M_{\star}$ is known, we use as default 
the stellar mass-halo mass relation for low-mass galaxies, as found in \citet{rea17}, 
including the reported confidence intervals, to assign $M_{200}$ to each galaxy.
\citet{rea17} fit the rotation curves of a `clean'
sample of isolated dwarf galaxies plus Carina dSph galaxy and Leo T,
and found a monotonic relation between $M_{\star}$ 
and $M_{200}$ with little scatter. As a test of the $\Lambda$CDM model, they compare this relation with the
$M_{\star}$-$M_{200}$ relation obtained from abundance matching technique, 
using the stellar mass function from Sloan Digital Sky Survey (SDSS), and found
good agreement. 
We caution, however, that the $M_{\star}$--$M_{200}$ relation in
dwarf galaxies is still an active field of research:  galaxy formation models 
predict values of $M_{200}$ above those derived from abundance matching with
field galaxies \citep[e.g.,][]{con18,for18}. In addition, many of the dSph galaxies 
in our sample are satellites that may have suffered tidal stripping. This could move
dSph galaxies off of the $M_{\star}$-$M_{200}$ relation of isolated dwarf galaxies,
probably leading to a value of $M_{200}$ below those derived from abundance matching
\citep[e.g.,][]{err18}. Given the uncertainties in the $M_{\star}$-$M_{200}$ relation,
we will explore below how $\tau_{\smalls \rm DF}$ depends on the assumed scatter 
in this relation.

Besides $M_{200}$, we use the concentration parameter $c$
to characterize the NFW profile of dark matter haloes.
We sample the concentration $c$ using the distribution function given
in \citet{sha21} (their figure 2).
These authors fit an NFW profile to field galaxies with stellar mass between $2\times
10^{7}M_{\odot}$ and $8\times 10^{7}M_{\odot}$ formed in the E-MOSAICS simulation.
The distribution function of the concentration has a median value of 
$c=10.5$, a value consistent with previous high-resolution cosmological simulations. 

For illustration, Figure \ref{fig:Fornax} shows the distribution 
function of $M_{200}$ and $r_{s}$ for a galaxy with the luminosity of Fornax. 
For our fiducial model, the halo mass peaks at a
value of $10^{10} M_{\odot}$, whereas $r_{s}$ peaks at $4$ kpc.
This halo mass is larger than the value adopted
by other authors \citep[e.g., $2.7\times 10^{9}M_{\odot}$ in][]{mea20}, 
but consistent with the mass estimate derived by \citet{rea19}, 
$(2.2\pm 0.7) \times 10^{10}M_{\odot}$, using abundance matching
\citep[see also][]{sha21}. 
On the other hand, our estimate of $r_{s}$ is larger than
the value estimated in \citet{mea20} ($2.1$ kpc), but smaller than the median
value ($6$ kpc) found in E-MOSAICS by \citet{sha21}. To illustrate the sensitivity
of the inspiral rate on the uncertainties in the $M_{\star}$-$M_{200}$
relation, 
Figure \ref{fig:Fornax} also shows $\tau_{\rm \smalls DF}$ of a GC with 
a mass of $2\times 10^{5}M_{\odot}$
(which is the median mass of Fornax GCs) at an initial radius of $1$ kpc
using the amount of scatter described above, but also when the 
amount of scatter in $M_{200}$ for a given stellar mass $M_{\star}$ has been tripled.
We see that it has little impact on the value of $\tau_{\rm \smalls DF}$.
Finally, we also show the corresponding distributions if, instead of the 
distribution function of the concentration given in \citet{sha21}, we sample the 
concentration using a normal distribution around $c=14$ with 
$\sigma_{c}=3$ \citep[e.g.,][]{dut14}. A larger concentration index implies 
a smaller $r_{s}$ and a slightly longer ($10$ percent) $\tau_{\rm \smalls DF}$.

It is likely that tidal forces and energetic outflows from bursty star formation
can kinematically heat up the dark matter at the central parts, forming cores.
Dwarf-dwarf mergers could also push the GCs towards more extended orbits.
By ignoring these effects, we are overestimating the rate of orbital decay of GCs
(see Section \ref{sec:discussion}).

A trial of the GC system is performed by
assigning either $38$ or $55$ GCs, depending if we use only the spectroscopically
confirmed GCs or the extended sample, to their respective galaxies. 
The mass of each GC, $M_{\smalls \rm GC}$, is computed from its magnitude, 
whose values are given in Table \ref{table:GC_properties}, and the mass-to-light 
ratio $(M/L)_{\smalls \rm GC}$.
The latter quantity, $(M/L)_{\smalls \rm GC}$, is randomly drawn from a uniform 
distribution between $1.5M_{\odot}/L_{\odot}$ and $2.5M_{\odot}/L_{\odot}$ 
\citep{mcl05}.

The time since GCs started orbital decay, $\tau_{\smalls \rm GC}$, corresponds to
their age $\tau_{\rm age}$ for in-situ GCs, but it is less than their age for accreted GCs. 
There is no a simple way to distinguish between in-situ and accreted GCs.
Although there are indications of late time mergers (e.g., Coleman et al.
2004), accretion of the GCs by dwarf-dwarf mergers are expected 
to occur mosty at redshifts $z\geq 2$ (corresponding to $\geq 9$ Gyr).
As a result, accreted GCs tend to be old, whereas GCs younger than $7$ Gyr are 
expected to be in-situ GCs. Therefore,  we will assume that the
five GCs with estimated ages between $6$ and $7$ Gyr in Table \ref{table:GC_properties}
are in-situ GCs, and we take  for $\tau_{\smalls \rm GC}$ the reported age values 
including their uncertainties.  For the remainder GCs, $\tau_{\smalls \rm GC}$
is randomly drawn from a uniform distribution between $6$ and $12$ Gyr.
In fact, all those GCs have $\tau_{\rm age}$ consistent with $12$ Gyr (within one
standard deviation), except one. Note also that $55\%$ of the GCs in our
sample has no age determination. Therefore, a range of $\tau_{\smalls \rm GC}$
between $6$ and $12$ Gyr seems reasonable. However, we also present results
assuming that all GCs are in-situ and evolve for the full age.

\begin{figure}
\includegraphics[width=77mm, height=168mm]{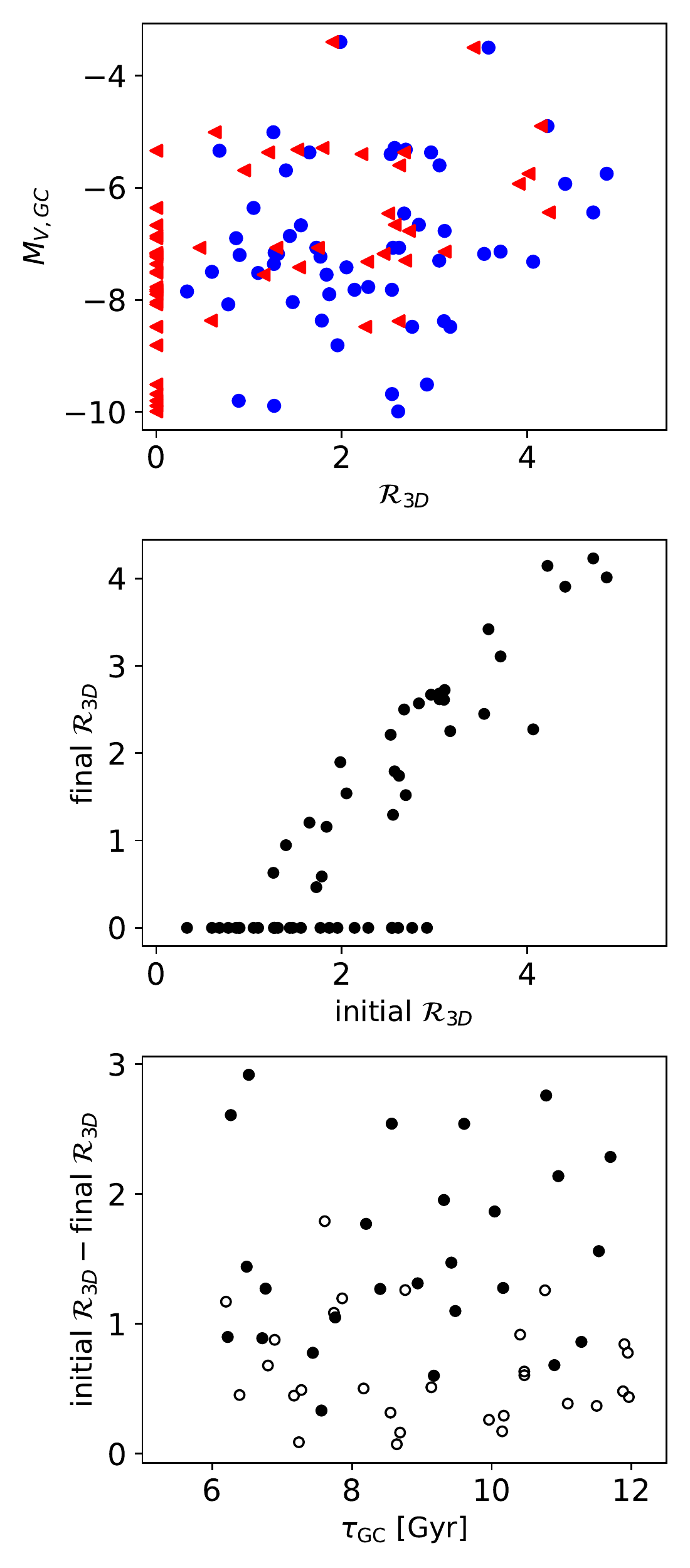}
 \caption{Evolution of the 3D distance from the host centre of GCs due to dynamical friction
in a realization with $\lambda=2$. The upper inset shows the starting radii (blue points) and
the final radii (red triangles) in the plane 
$\mathcal{R}_{3\smalls \rm D}$--$M_{\smalls V,\rm GC}$. The middle inset displays
the final versus the starting values of $\mathcal{R}_{3\smalls \rm D}$ in the same
realization. The lower panel shows the change in the orbital radius versus
$\tau_{\smalls \rm GC}$. Solid dots mark those GCs that have sunk to the
centre. 
}
\label{fig:decay_orbits}
\end{figure}

Note that we are not including mass loss of GCs in our models. If GCs were more massive
in the past, the dynamical friction timescale would become shorter
\citep[e.g.,][]{amo17}. Therefore,
our assumption of constant $M_{\rm \smalls GC}$ is underestimating dynamical
friction.
The problem of the survival of low-mass GCs that lie in the central regions 
(at least in projection) against tidal forces, such as Eridanus II GC or DDO 216-A1, 
is an interesting issue that may provide additional constraints to the distribution
of dark matter \citep[e.g.,][]{amo17,con18,lea20}.

In the next section, we will assume a simple distribution of the starting GC 
distances and compare the expected present-day radial distribution of GCs
with the observed one.
In Section \ref{sec:past_future}, we take the current projected distances of the GCs and use 
Equation (\ref{eq:final_radius_df}) to infer their past and future positions.

\begin{figure*}
\hspace{-0.2cm}
\includegraphics[width=169mm, height=214mm]{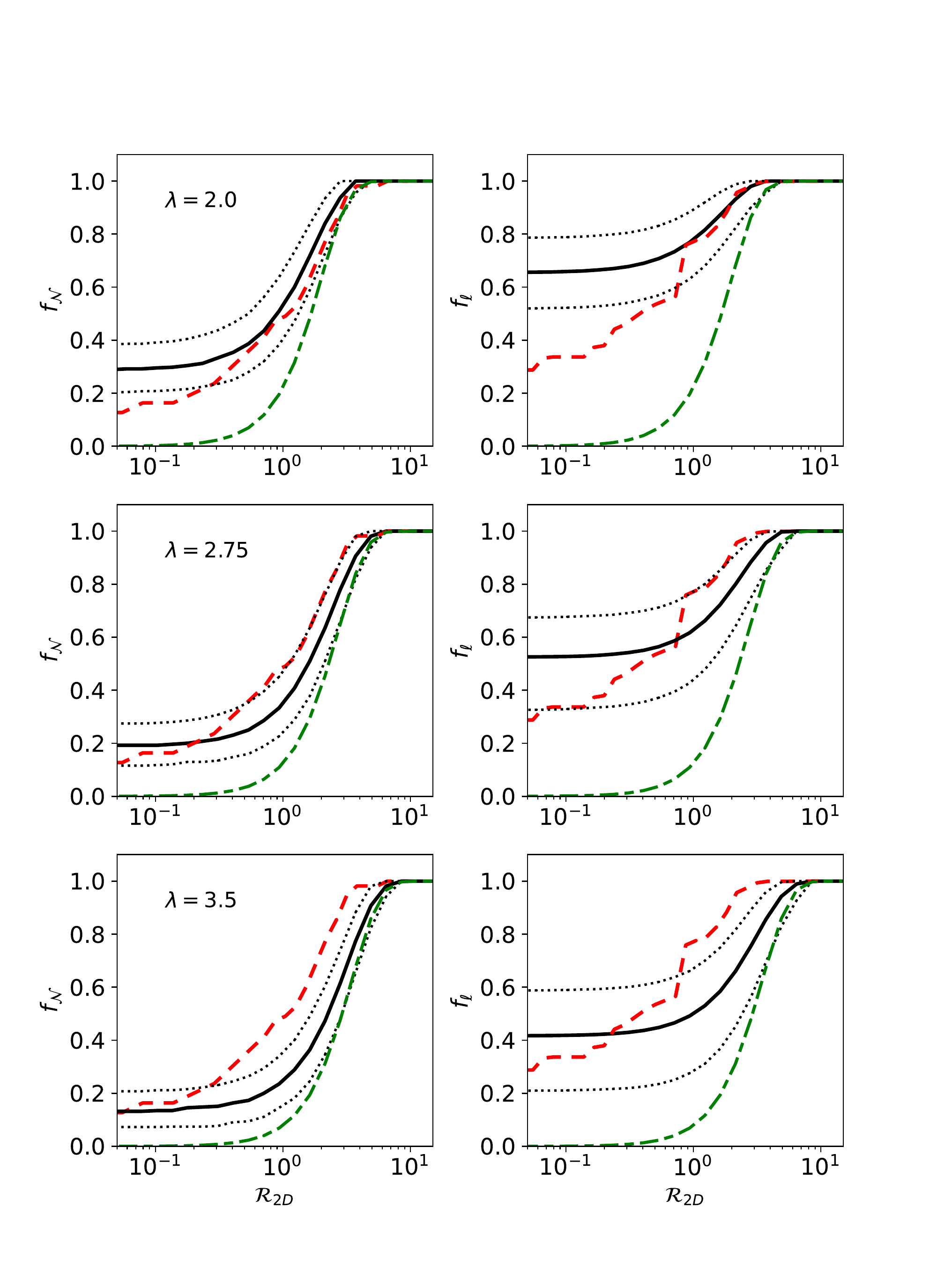}  
 \caption{Fraction of GCs (left panels) and fraction of specific luminosity (right panels) enclosed within a circle of radius $\mathcal{R}_{2\smalls \rm D}$ from the centre
of the galaxies, for three values of $\lambda$. The red dashed
lines show the observed distributions in the extended sample. The black curves are the 
predicted distributions in our Monte Carlo model at the present day
(solid lines indicate the median
value and the dotted lines encompass the $95.4$ percent of the Monte Carlo
simulations). The green dashed lines show the initial distributions.
}
\centering
\label{fig:cumulative_diffRadius}
\end{figure*}

\begin{figure*}
\hspace*{-1.3cm}\includegraphics[width=216mm, height=63mm]{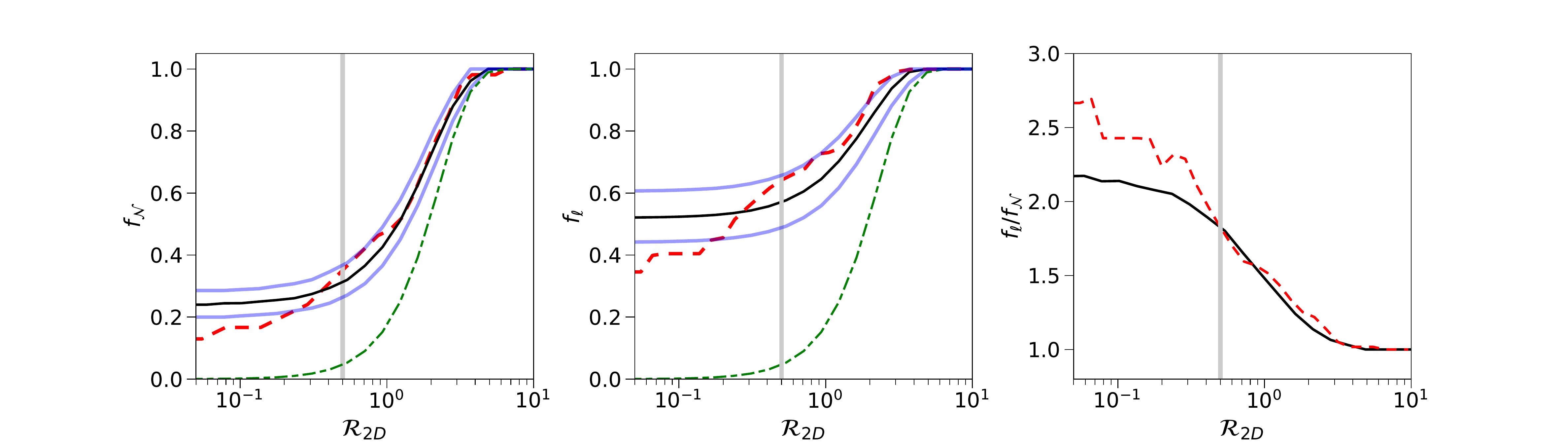}  
 \caption{Radial dependence of $f_{\smalls \mathcal{N}}$ (left panel), $f_{\ell}$ 
(midlle panel) and $f_{\ell}/f_{\smalls \mathcal{N}}$ (right panel), excluding 
KK 221 2-966. The black, red and green lines
have the same meaning as in Figure \ref{fig:cumulative_diffRadius}.
The lower blue lines and the upper blue lines indicate the $16$th and $84$th percentiles,
respectively. In these models, we assume $\lambda=2.3$.
The vertical lines mark the radial distance $\mathcal{R}_{2\smalls \rm D}=0.5$.
}
\centering
\label{fig:without_1GCs}
\end{figure*}

\begin{figure*}
\hspace*{-1.3cm}\includegraphics[width=216mm, height=63mm]{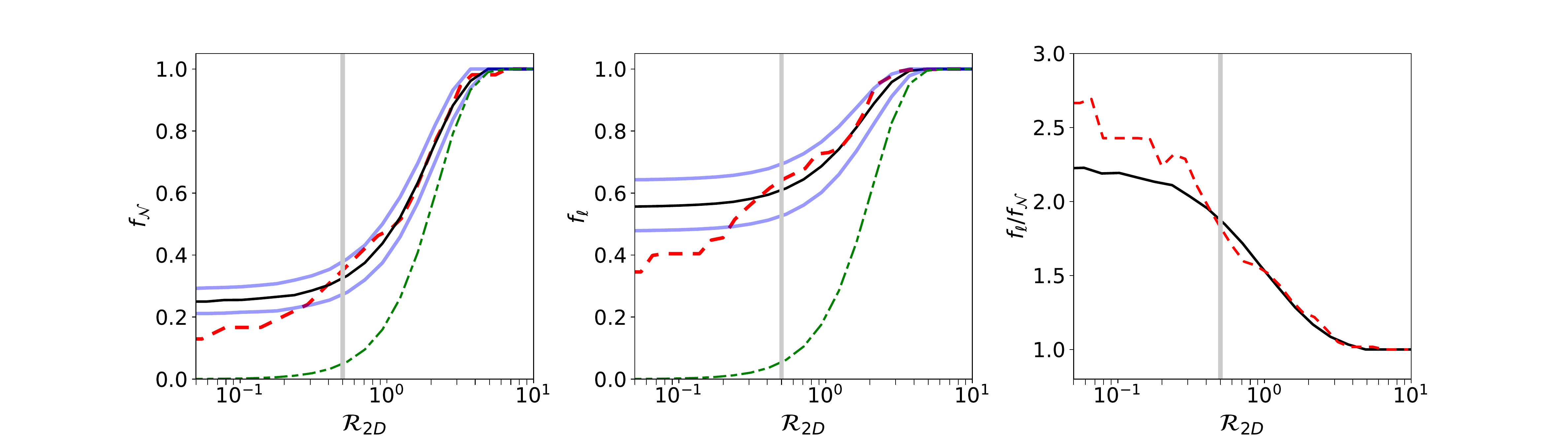}  
 \caption{Same as Figure \ref{fig:without_1GCs} but using $\lambda=2.1$
for GCs with $\ell >0.005$ and $\lambda=2.4$ for GCs having $\ell\leq 0.005$
(excluding  KK 221 2-966).
}
\raggedright
\label{fig:initially_segregated}
\end{figure*}

\begin{figure}
\hspace*{-0.7cm}
\includegraphics[width=102mm, height=86mm]{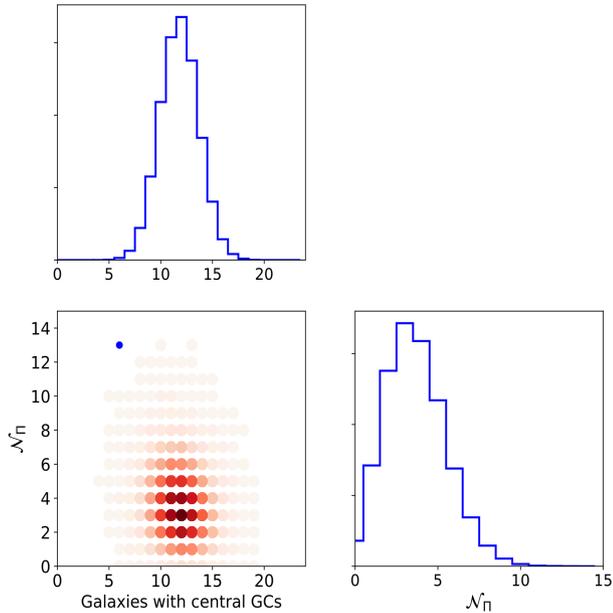}  
 \caption{Corner plot showing the correlation and distributions of the number of
galaxies with a central GC and $\mathcal{N}_{\Pi}$ (the number of GCs inside 
$\mathcal{R}_{2\smalls \rm D}=0.5$, but excluding central ones), 
at the present epoch, in a model with
$\lambda=2.3$. The blue dot indicates the data point in that plane.
}
\centering
\label{fig:central_periphery}
\end{figure}

\begin{figure}
\hspace*{-0.48cm}
\includegraphics[width=99mm, height=47mm]{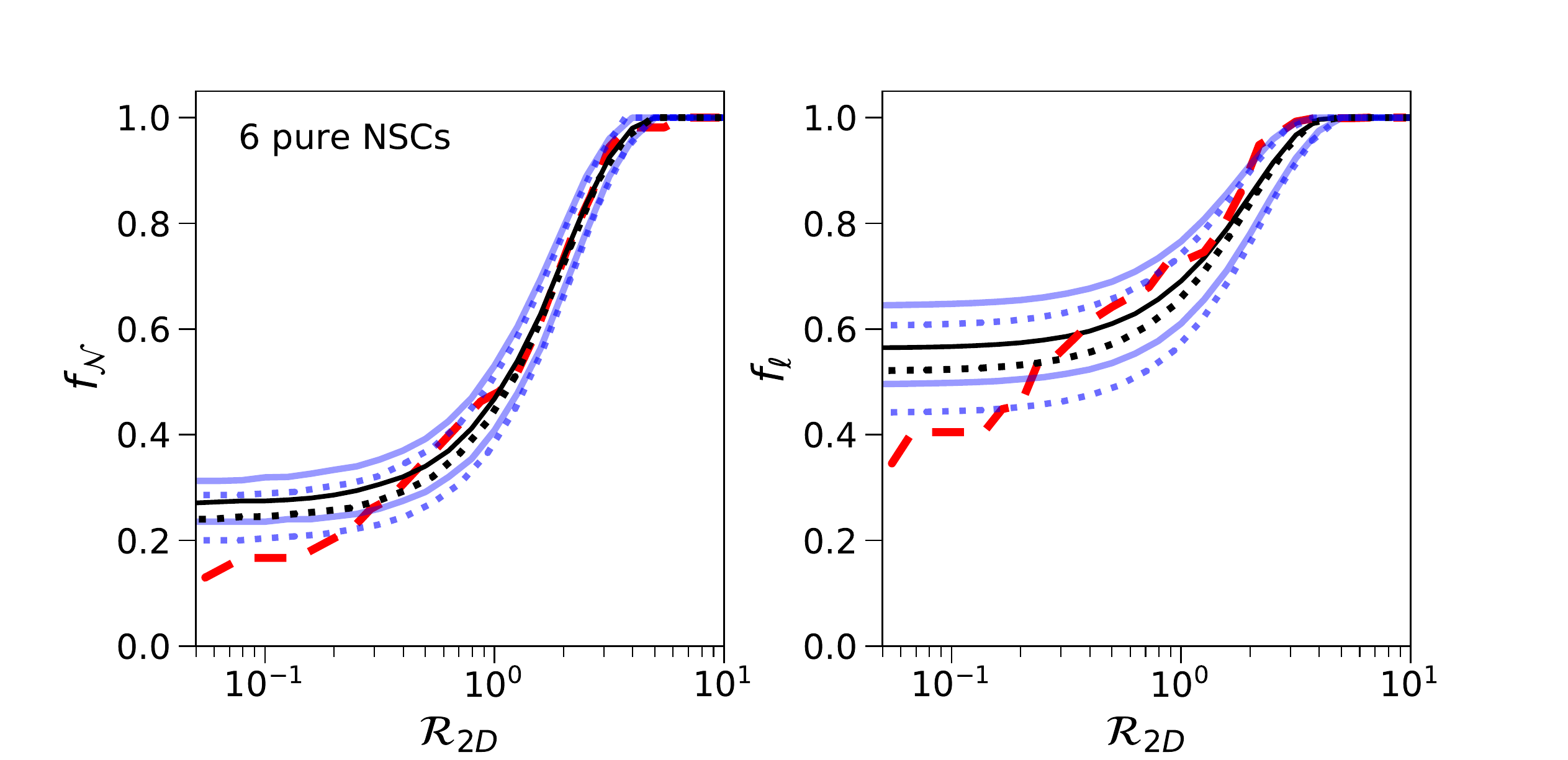}  
 \caption{Radial dependence of $f_{\smalls \mathcal{N}}$ and $f_{\ell}$ in a model
where the six NSCs are assumed to have formed in situ at the galaxy centre, i.e. 
they are pure NSCs (solid lines). The starting radial distances of the other GCs 
are given by Equation (\ref{eq:NGC_Rinit}) with scale $\lambda=2.3$.
To facilitate comparison, the corresponding curves for the fiducial model (without
assuming pure NSCs) are also shown (dotted lines). Line colours
are the same as Figures \ref{fig:without_1GCs} and \ref{fig:initially_segregated}.}
\centering
\label{fig:6NSC_no_stalling}
\end{figure}

\begin{figure}
\hspace*{-0.5cm}
\includegraphics[width=99mm, height=47mm]{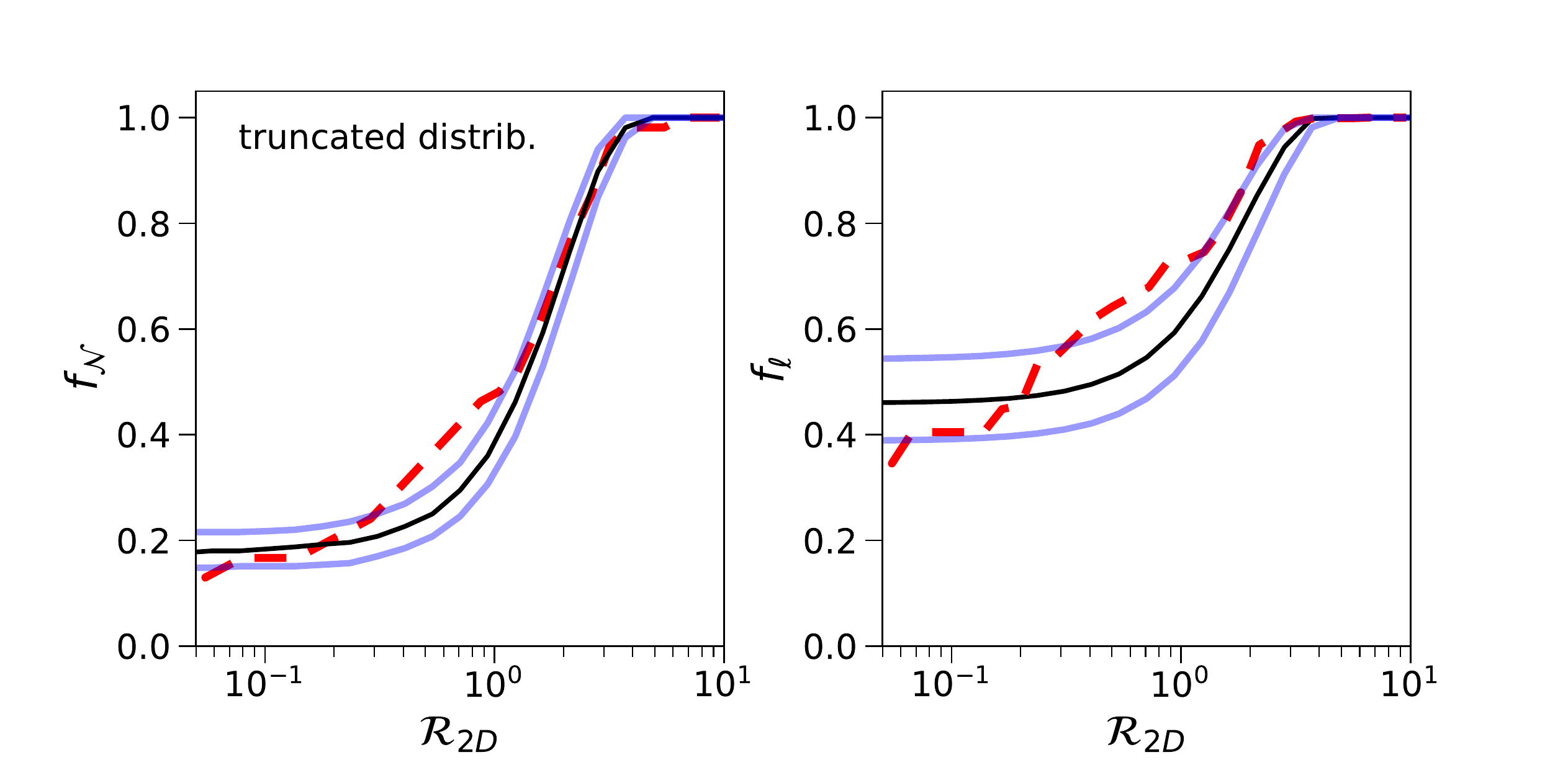}  
 \caption{$f_{\smalls \mathcal{N}}$ and $f_{\ell}$ versus $\mathcal{R}_{2\smalls D}$
 in a model
where the distribution of starting radial distances is given by Equation (\ref{eq:NGC_Rinit})
with scale $\lambda=2.3$, but truncated at an inner radius 
$\mathcal{R}_{2\smalls \rm D}=1.5$ and at an outer truncation radius $\mathcal{R}_{2\smalls \rm D}=4.6$. 
}
\centering
\label{fig:truncated_Gaussian}
\end{figure}

\section{What to expect if the initial radial distribution is Gaussian. A heuristic approach}
\label{sec:modelA}
\subsection{Fiducial model}
It is convenient to define $\mathcal{R}_{3\smalls \rm D}$ as the $3$D distance
to the host centre in units of the galaxy effective radius.
As the simplest starting assumption, we consider in this section that the orbits 
of the surviving GCs are almost circular, i.e. $\tilde{e}=0$, and 
the probability distribution function (PDF) of ${\mathcal{R}}_{3\smalls \rm D}$,
that is the probability that a GC has a starting distance 
${\mathcal{R}}_{3\smalls \rm D}$, is 
\begin{eqnarray}
{\mathcal{N}}_{\rm GC}^{(i)} ({\mathcal{R}}_{3\smalls \rm D})= 
4\pi^{-1/2} \lambda^{-3} \mathcal{R}_{3\smalls \rm D}^{2}
\exp\left(-\left[\frac{\mathcal{R}_{3\smalls \rm D}}{\lambda}\right]^{2}\right),
\label{eq:NGC_Rinit}
\end{eqnarray}
where $\lambda$ is a dimensionless free parameter. 
The superscript $(i)$ indicates that the distribution is the initial one. The above distribution
indicates that the volume probability density for a GC to start at 
$\mathcal{R}_{3\smalls \rm D}$ is Gaussian.
The corresponding PDF of $\mathcal{R}_{2\smalls \rm D}$ is
\begin{eqnarray}
{\mathcal{N}}_{\rm GC}^{(i)} ({\mathcal{R}}_{2\smalls \rm D})= 2\lambda^{-2} 
\mathcal{R}_{2\smalls \rm D} \exp\left(-\left[\frac{\mathcal{R}_{2\smalls \rm D}}{\lambda}\right]^{2}\right).
\label{eq:initial_pdf_2D}
\end{eqnarray}
Thus, the effective radius of the GC system as a whole is $\sqrt{\ln 2} \lambda R_{e}$.
In this simple model, the starting position of the GCs is correlated neither with its mass,
i.e. there is no mass segregation initially, nor with its age.

The way to set the initial radius of GCs in the Monte Carlo model is as follows.
In each trial, we draw the initial $\mathcal{R}_{3\smalls \rm D}$ from the PDF given in
Equation (\ref{eq:NGC_Rinit}). We next convert them into physical units by sampling
the effective radius of each galaxy $R_{e}$ from a normal distribution around the
observed mean value and the $1\sigma$ uncertainty, which are given in Table \ref{table:dSph_properties}.
Following the method described in the previous section, we can derive the 
radial distributions of $\mathcal{R}_{3\smalls \rm D}$, $\mathcal{R}_{2\smalls \rm D}$, 
and $\ell$ at the present day.

Figure \ref{fig:decay_orbits} shows the starting and the final distances 
$\mathcal{R}_{3\smalls \rm D}$ of the GCs, in a realization of 
$55$ GCs with $\lambda=2$. It is clear that GCs far 
away from the centre of the host galaxy undergo comparatively less orbital
decay because the density of dark matter is comparatively small. Inner GCs are
expected to experience more friction. In that particular realization, 
we can see that most of the GCs that starts within $\sim 1.5R_{e}$ 
sink to the centre of their host 
galaxies, regardless their magnitude $M_{\smalls V,\rm GC}$.  For those GCs starting
at distances larger than $2R_{e}$, only the most luminous and hence the
most massive GCs suffer an appreciable change in their orbital radii. In a large
fraction of the trials, the five most luminous GCs sink to the centre. From the
lower panel, we see that the values of $\tau_{\rm \smalls GC}$ of those GCs 
that reach the centre span across the whole $\tau_{\rm \smalls GC}$ range.

Figure \ref{fig:cumulative_diffRadius} shows the starting and the present-day 
cumulative distributions of projected distances, denoted by 
$f_{\smalls \mathcal{N}}(\mathcal{R}_{2\smalls \rm D})$, for 
$\lambda=2.0, 2.75$, and $3.5$.  We also show 
$f_{\ell}(\mathcal{R}_{2\smalls \rm D})$,
which is defined as the fraction of specific luminosity contained inside a 
projected radius $\mathcal{R}_{2\smalls \rm D}$.
At each radius ${\mathcal R}_{2\smalls \rm D}$, we provide
the median value and $2.3$th and $97.7$th percentiles of the 
distributions after $1.2\times 10^{5}$ realizations. We also plot 
$f_{\smalls \mathcal{N}}$ and$f_{\smalls \ell}$ as derived from the observational data.

For $\lambda=2$, approximately $30\%$ of the GCs sink to
the host galaxy centre at the present day (top left panel). 
These central GCs contain more than $60\%$ of the total specific 
luminosity (top right panel). Due to the inspiraling of GCs,
the effective radius of the GC system 
(i.e. the radius at which $f_{\smalls \mathcal{N}}=0.5$),
changes from $1.66R_{e}$ initially to $0.92R_{e}$ at the present day.
As expected, the number of GCs that sink to the host centre decreases with $\lambda$.
In order to account for the observed fraction of GCs at
$\mathcal{R}_{2\smalls \rm D}=0.1$, a value of $\lambda\simeq 3.5$ is required. 
However, this model underpredicts the number of GCs over a large radial 
range. For $\lambda=2.75$, $f_{\ell}$ at $\mathcal{R}_{2\smalls \rm D}=0.1$ is larger than the observed value in $97.7$ percent
of the trials (middle right panel in Fig. \ref{fig:cumulative_diffRadius}).
At $1<\mathcal{R}_{2\smalls \rm D}<4$, however, more than $97.7$ percent 
of the trials give smaller values for $f_{\smalls \mathcal{N}}$ and $f_{\ell}$
than the observed ones.

We noticed that the GC KK 221 2-966 sinks to the centre of the host 
galaxy in a significant fraction of the trials if $\lambda \leq 3.0$. However, 
the current 2D distance of this GC is $0.85R_{e}$. 
Indeed, the jump in $f_{\ell}$ at 
$\mathcal{R}_{2\smalls \rm D}=0.85$ visible in the observed distribution of the 
extended sample (red dashed curves in the right panels of 
Figure \ref{fig:cumulative_diffRadius}) 
is due to the contribution to $\ell$ by KK 221 2-966.  In fact, the form 
of $f_{\ell}$ versus $\mathcal{R}_{2\smalls \rm D}$
is dominated by the radial distance of the most luminous GCs.
This produces that $f_{\ell}$ is generally less smooth than $f_{\smalls \mathcal{N}}$. 
Therefore, we decided to explore how the radial distributions 
change when this GC is excluded from the analysis and it is studied separately
(see Section \ref{sec:starting_dist}).

Figure \ref{fig:without_1GCs} shows the radial distributions derived
from the observational data,
together with the expected distributions for a model with $\lambda=2.3$,
once KK 221 2-966 is excluded. To facilitate the discussion, let us first focus
on the region $\mathcal{R}_{2\smalls \rm D}\gtrsim 0.5$. We see that the
predicted curve of $f_{\smalls \mathcal{N}}$ follows quite well the
distribution as derived from the data. Between
$\mathcal{R}_{2\smalls \rm D}= 0.5$ and $\mathcal{R}_{2\smalls \rm D}= 1$,
the observed values of $f_{\smalls \mathcal{N}}$ and $f_{\ell}$ are larger than
the predicted median values, but they lie within the $68$th percentile.
The enhancements in the observed values of $f_{\smalls \mathcal{N}}$ and $f_{\ell}$
are correlated, so that the ratio $f_{\ell}/f_{\smalls \mathcal{N}}$ follows the ratio 
expected in that model. 
In the case under consideration, the curve $f_{\ell}/f_{\smalls \mathcal{N}}$ 
as derived from the data is sligthly above the predicted curve. 
In addition, the correlation coefficient between $\ell$ and 
$\mathcal{R}_{2\smalls \rm D}$, denoted by $\hat{\rho}_{s}$, is $-0.45\pm 0.12$ 
for the observed distribution, whereas it is $-0.31\pm 0.11$ in our 
Monte Carlo model 
with $\lambda=2.3$ \footnote{$\hat{\rho}_{s}$ depends on the adopted value 
of $\lambda$. In particular, we obtain that $\hat{\rho}_{s}=-0.34\pm 0.11$ for
$\lambda=2$, and $-0.28\pm 0.11$ for $\lambda=2.75$.}.
Although these values of $\hat{\rho}_{s}$ are consistent  
at $1\sigma$ level, $\hat{\rho}_{s}$ can be enhanced 
by introducing some degree of primordial mass segregation.
For instance, if we split the population of GCs into two halves; those
having $\ell >0.005$ and those with $\ell <0.005$, and assign
a scalelength parameter of $\lambda=2.1$ for the first group and 
$\lambda=2.4$ for second group, the initial value of $\hat{\rho}_{s}$
is $-0.12\pm 0.13$, and its final value is $-0.39\pm 0.11$,
which is closer to the real value. The resulting distributions in
this two-Gaussian model are shown in Figure \ref{fig:initially_segregated}.
We find that the global properties of the observed radial distribution 
of GCs at distances $\mathcal{R}_{2\smalls \rm D}\gtrsim 0.5$, 
once the massive GC KK 221 2-966 is excluded, can be reasonably accounted for 
in a two-Gaussian model.

\begin{figure}
\hspace{-0.8cm}
\includegraphics[width=100mm, height=48mm]{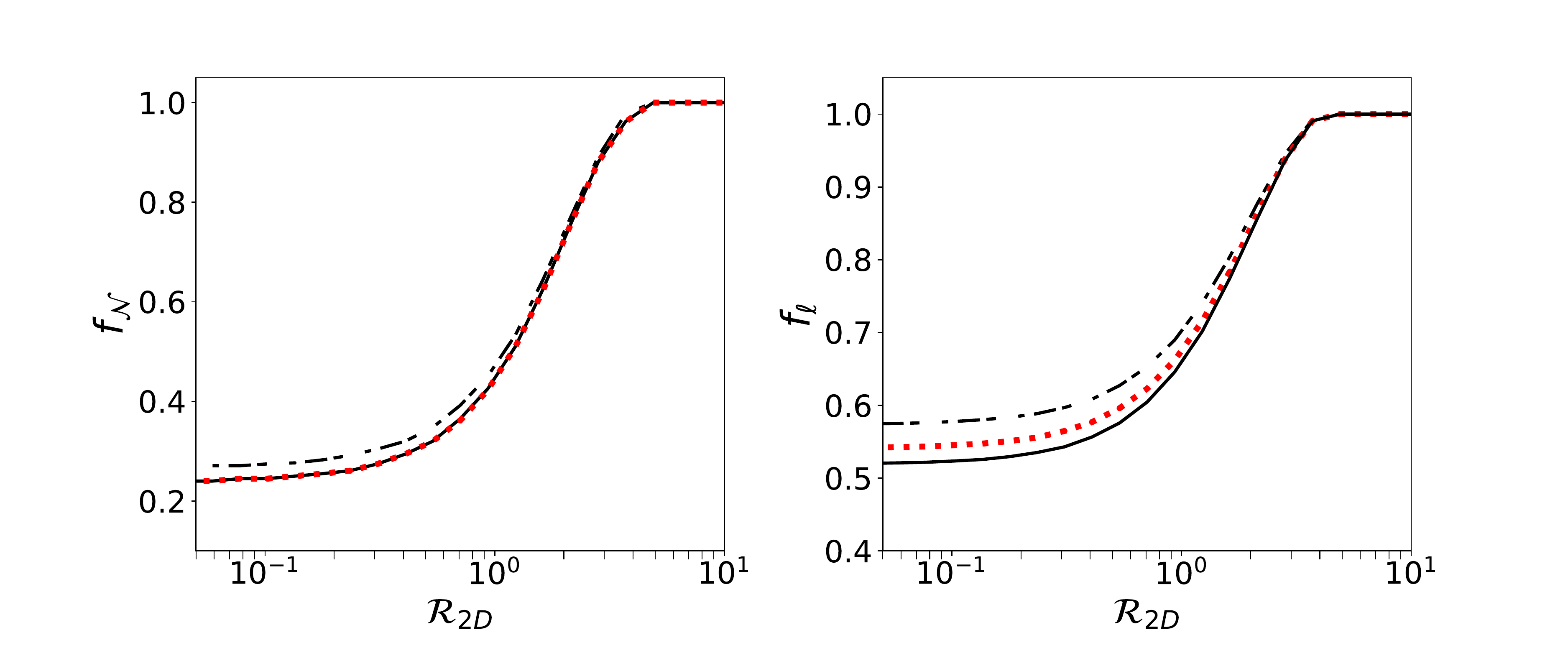}  
 \caption{Distributions $f_{\smalls \mathcal{N}}$ and $f_{\ell}$,
adopting different recipes for $\tau_{\smalls \rm GC}$. Black solid lines correspond to
our fiducial model described in Section \ref{sec:MC_model}.
Black dash-dotted lines are the results when adopting $\tau_{\rm \smalls GC}=\tau_{\smalls
\rm age}$, and a uniform distribution over the interval $[6,13.5]$ Gyr
for those GCs that have no estimates of their ages.  Finally, red dotted curves correspond
to the results assuming $\tau_{\rm \smalls GC}=\tau_{\smalls\rm age}$ for those having
age estimates, and $\tau_{\rm \smalls GC}=7$ Gyr for those that do not have age estimates.
In all these models, we assume $\lambda=2.3$.
}
\centering
\label{fig:full_age}
\end{figure}

Consider now the radial GC distribution at 
$\mathcal{R}_{2\smalls \rm D}\lesssim 0.5$. The predicted functions
$f_{\smalls \mathcal{N}}$ and $f_{\ell}$ at $\mathcal{R}_{2\smalls \rm D}<0.5$ 
for $\lambda=2.3$ are shallower than the observed distributions 
(Figure \ref{fig:without_1GCs}). The predicted values of $f_{\smalls \mathcal{N}}(0)$ 
and $f_{\ell}(0)$ for the model with $\lambda=2.3$ are higher
than the observed values.  Note that $f_{\smalls \mathcal{N}}(0)$ is the
fraction of central GCs. In the data sample, six galaxies have a central GC. However,
in only $0.2\%$ of the trials in a model with $\lambda=2.3$, six
galaxies (or less) have a central GC (see Figure \ref{fig:central_periphery}). 
In fact, the expected number of galaxies with a central GC is $12\pm 2$ 
(excluding KK 221 2-966 in the analysis).

\begin{figure}
\hspace*{-0.7cm}
\includegraphics[width=100mm, height=48mm]{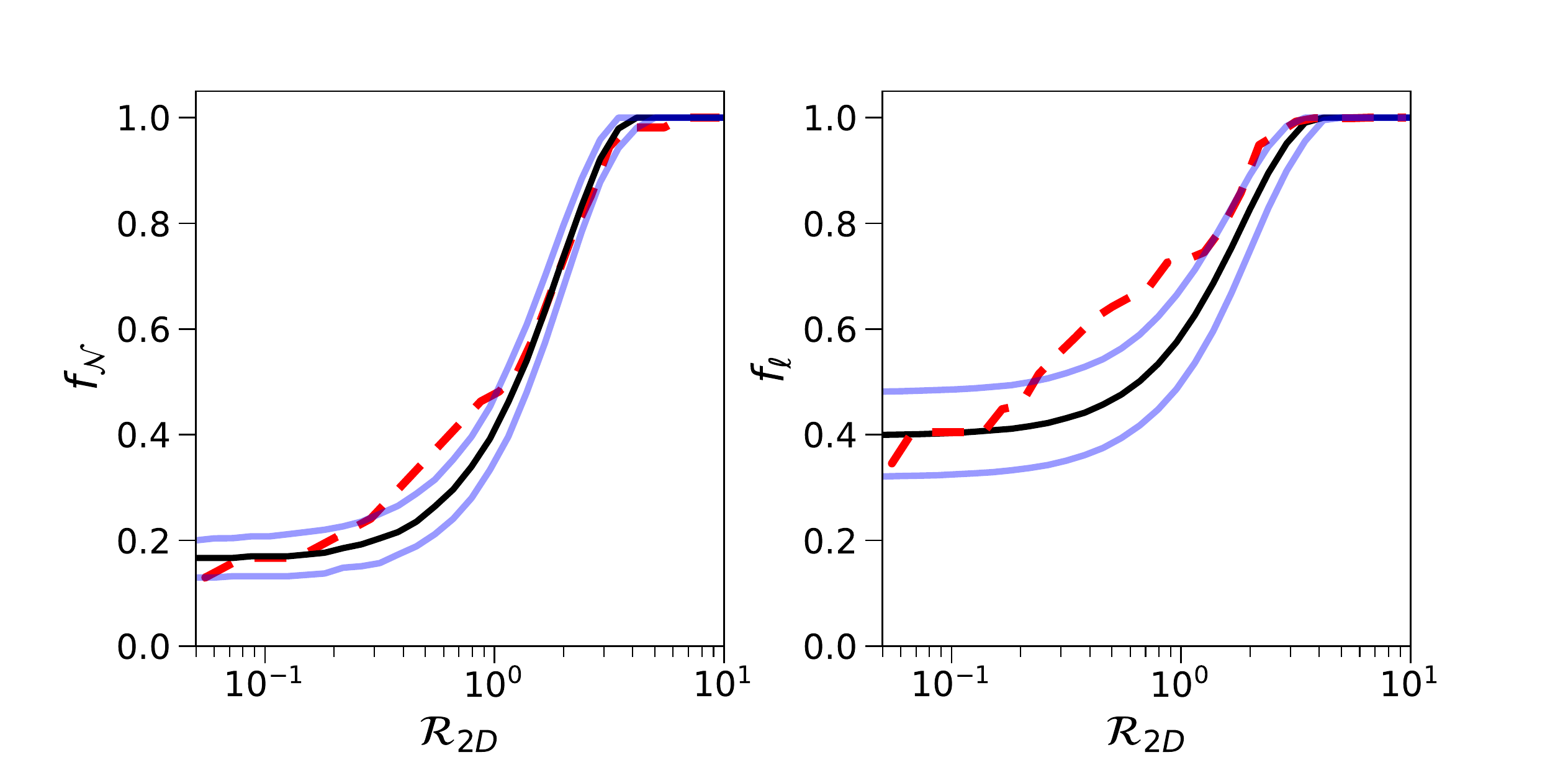}  
 \caption{Radial profiles of $f_{\smalls \mathcal{N}}$ and $f_{\ell}$ in a model
where the value of $\tau_{\rm \smalls DF}$, given in
Equation (\ref{eq:tau_DF_Arca}), is increased by a factor of $2$. 
The distribution of starting distances is given by Equation (\ref{eq:NGC_Rinit}) with
$\lambda=2.0$. Line colours and style codes 
are the same as those used in Figures \ref{fig:without_1GCs}, \ref{fig:initially_segregated},
\ref{fig:6NSC_no_stalling} and \ref{fig:truncated_Gaussian}.
}
\centering
\label{fig:twice_ArcaS}
\end{figure}

We also observe a discrepancy between the observed and the predicted slope of 
$f_{\smalls \mathcal{N}}$ at $\mathcal{R}_{2\smalls \rm D}\lesssim 0.5$,
which indicates a discrepancy in the number density of GCs. To make a quantitative
comparison, it is useful to define $\mathcal{N}_{\Pi}$ as the number of GCs with
projected radii $0<\mathcal{R}_{2\smalls \rm D}<0.5$ (excluding the nuclei). 
More formally, 
$\mathcal{N}_{\Pi}=[f_{\smalls \mathcal{N}}(0.5)-f_{\smalls \mathcal{N}}(0)]N_{\smalls \rm GC}$,
where $N_{\smalls \rm GC}$ is the total number of GCs.
In our extended sample of GCs,
the value is $\mathcal{N}_{\Pi}=13$. However,
the model with $\lambda=2.3$ predicts $\mathcal{N}_{\Pi}=4\substack{+1\\-2}$ 
because there GCs spend little time given that dynamical friction is strong. 
Figure \ref{fig:central_periphery} shows the probability distribution 
of $\mathcal{N}_{\Pi}$ and the correlation between $\mathcal{N}_{\Pi}$ and 
the number of galaxies hosting a NSC.
In this model, the probability that $\mathcal {N}_{\Pi}\geq 13$
is $\sim 8\times 10^{-5}$.
This discrepancy between predictions and observed values is also obtained when
realizations of the $38$ confirmed GCs are carried out\footnote{In the sample
of $38$ confirmed GCs, $\mathcal{N}_{\Pi}=10$.  In our Monte Carlo models,
the probability that $\mathcal{N}_{\Pi}\geq 10$  
is $\sim 2\times 10^{-4}$. On the other hand, we find that six galaxies (or less) host a
central GCs in $0.5\%$ of the trials.}.

In the two-Gaussian model, which includes some degree of primordial mass
segregation, the discrepancy between the predicted distributions and the distributions
derived from observations at $\mathcal{R}_{\smalls 2D}<0.5$ increases, especially for 
$f_{\ell}$. The reason is that in this model the most luminous GCs are initially more 
spatially concentrated, so that the GCs that arrive to radii $\mathcal{R}_{\smalls 2D}<0.5$ are, on average, more luminous than in a model without primordial mass segregation.

In order to find possible explanations for the offset in the
value of $\mathcal{N}_{\Pi}$ from the model distribution, it is worth
bearing in mind that, in our extended sample of GCs, $12$ galaxies contribute to 
$\mathcal{N}_{\Pi}$. Only IKN has two GCs at $0<\mathcal{R}_{\smalls 2D}<0.5$.
Therefore, none of the galaxies predominantly contributes to the
value of $\mathcal{N}_{\Pi}$.
On the other hand, in this subsample of $12$ galaxies 
that contribute to $\mathcal{N}_{\Pi}$, one galaxy (KK 197) has also a NSC. 
Given that the fraction of galaxies hosting a GC is $6/55$ in our extended sample,
about one galaxy is indeed expected to have a NSC in a sample of $12$ galaxies, 
as long as there is no correlation between
hosting a GC at $0<\mathcal{R}_{\smalls 2D}<0.5$ and the possession of a NSC.

\begin{figure}
\hspace*{0.5cm}\includegraphics[width=75mm, height=105mm]{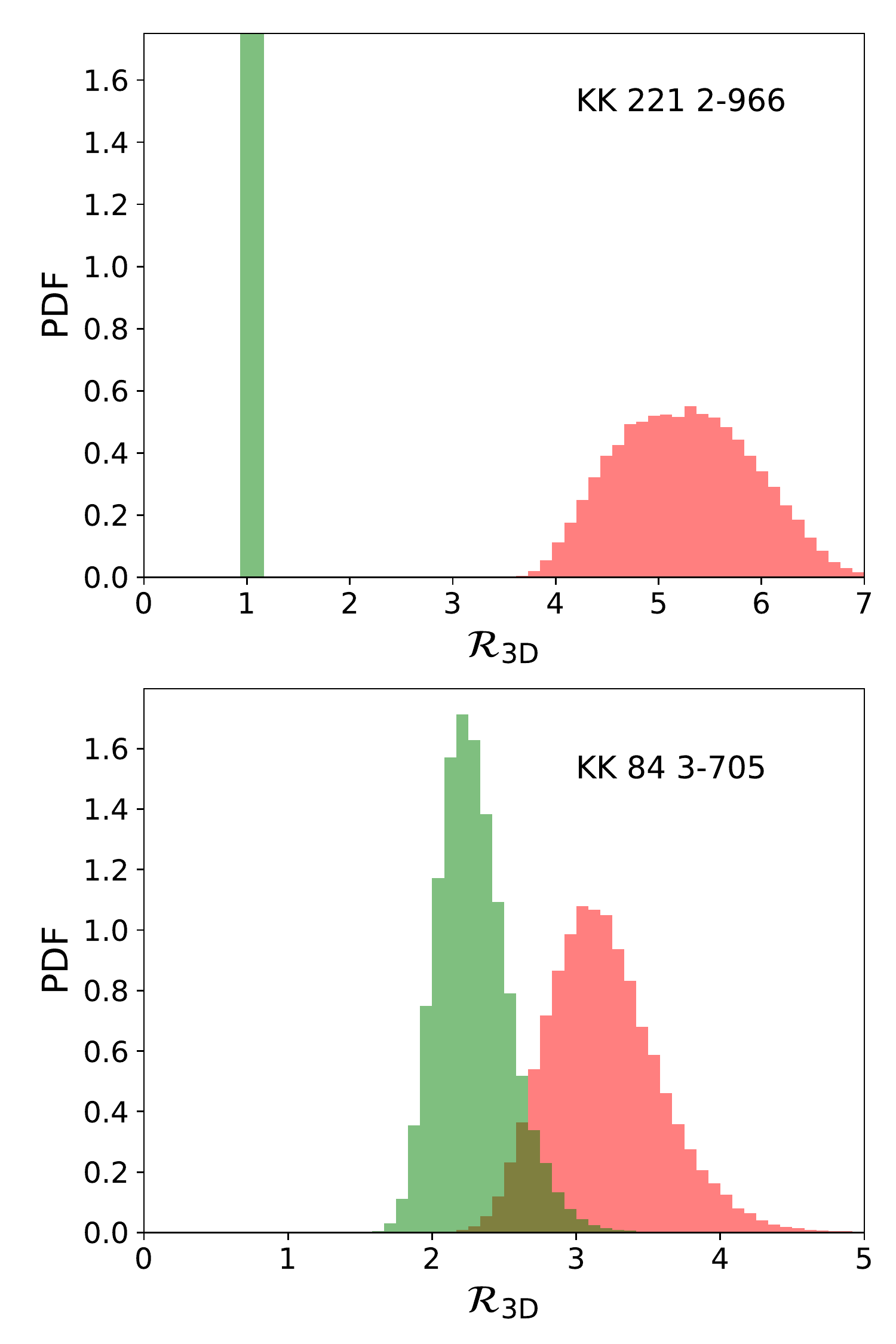}  
 \caption{Distribution of the present (green histograms) and
starting (red histograms) 3D distances for KK 221 2-966 (top panel) and
KK 84 3-705 (bottom panel).
}
\raggedright
\label{fig:initial_radius_2GCs}
\end{figure}

\begin{figure}
\hspace*{0.0cm}\includegraphics[width=90mm, height=60mm]{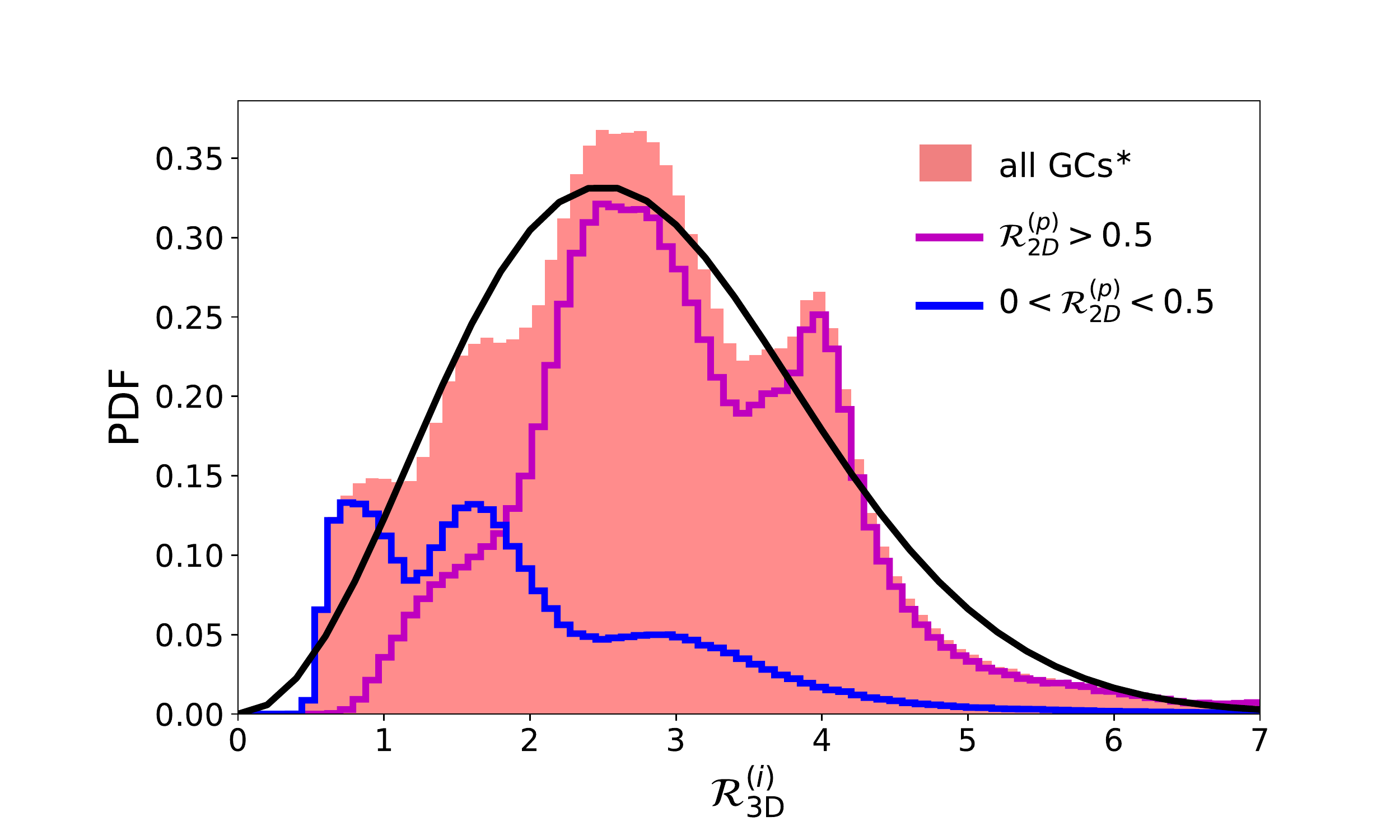}  
 \caption{Histograms of the distribution of the {\it starting} orbital radii of GCs, 
$\mathcal{N}_{\smalls \rm GC}^{(i)}$, in the full sample excluding central GCs (orange),
together with the contributions from GCs having $0<\mathcal{R}_{\smalls 2D}^{(p)}<0.5$ (blue histogram)
and from GCs with $\mathcal{R}_{\smalls 2D}^{(p)}>0.5$ (magenta).
The solid black line corresponds to
the graph of the function given in Equation (\ref{eq:NGC_Rinit}) with $\lambda=2.5$.
}
\raggedright
\label{fig:initial_radius_allGCs}
\end{figure}

\subsection{Variations of the fiducial model}
\label{sec:variations}

As said in Section \ref{sec:sample}, six GCs in our sample reside at the centre
of their host galaxies (And XXV Gep I, KKs 58-NSC, KK 197-02, KK 211 3-149,
ESO 269-66-03 and KK 84 3-830).
It is plausible that some of these central GCs did not inspiral towards
the centre but they were formed in situ at the centre of the host galaxy. 
From Table \ref{table:GC_properties}, we see that four 
of these GCs are more 
luminous than magnitude $-9.5$ (see also Figure \ref{fig:l_vs_R2D}).
In the in-situ formation scenario, an explanation for the NSCs being more luminous 
than non-central GCs could be that 
NSCs grow their stellar mass by sustained star formation due to the large amounts 
of gas available at the centre. We do not know what NSCs in our sample were
formed at the centre and what NSCs spiralled into the centre. 
We have computed $f_{\smalls \mathcal{N}}$ and
$f_{\ell}$ in the extreme situation where the six abovementioned NSCs are assumed 
to be formed in situ at the galaxy centres.  For these GCs we set their initial
orbital radius to zero and proceed following the Monte Carlo scheme described in
Section \ref{sec:MC_model}. The initial orbital radii of non-central GCs are drawn
from the distribution given in Equation (\ref{eq:NGC_Rinit}).
Figure \ref{fig:6NSC_no_stalling} shows the predicted distributions for the 
fiducial value of $\lambda$.
As expected, the predicted value of $f_{\smalls \mathcal{N}}(0)$
becomes larger (more discrepant with the observed values)
if some of the central GCs are pure NSCs.

The large fraction of galaxies with a central GC predicted in a model with $\lambda=2.3$
can be alleviated by introducing inner and outer truncation radii to the starting 
spatial distribution of GCs given in Equation (\ref{eq:NGC_Rinit}). 
For instance, Figure \ref{fig:truncated_Gaussian} shows the radial distributions for
$\lambda=2.3$, but assuming that no GCs were born
or started its orbital decay inside $1.5R_{e}$ and beyond $4.6R_{e}$.
In that model, the probability of having six (or less) galaxies with a central
GC is $5\%$. However, the predicted value of $\mathcal{N}_{\Pi}$ is still 
low as compared to the value derived from the observational data.
Moreover, it is hard to explain the origin of the internal
truncation on the basis that stellar cluster formation 
should have been more likely in the inner regions ($<1.5 R_{e}$) of the galaxy,
where a larger gas reservoir has been available to promote GC formation.

The difference $f_{\smalls \mathcal{N}}(0.5)-f_{\smalls \mathcal{N}}(0)$
 is rather insensitive to the value of $\lambda$,
as can be seen in Figure \ref{fig:cumulative_diffRadius},
implying that $\mathcal{N}_{\Pi}$ is also insensitive. It is worth to check
if $\mathcal{N}_{\Pi}$ is sensitive to our recipe for $\tau_{\smalls \rm GC}$.
In our fiducial model, we consider that some of the GCs with $\tau_{\rm age}>7$
Gyr could be potentially `ex-situ' GCs. For them, we assumed a uniform distribution
of $\tau_{\rm \smalls GC}$ between $6$ and $12$ Gyr.
In order to quantify the impact of this assumption, Figure \ref{fig:full_age} shows
$f_{\smalls \mathcal{N}}$ and $f_{\ell}$ when all the GCs are treated as in-situ GCs.
More specifically, we take $\tau_{\rm \smalls GC}=\tau_{\rm age}$ for all the GCs
having age estimates as given in Table \ref{table:GC_properties}. For those GCs with no age 
measurements, we explore two scenarios: (i) all of them have
$\tau_{\rm \smalls GC}=7$ Gyr and (ii) $\tau_{\rm \smalls GC}$
follows a uniform probability distribution in the range $[6,13.5]$ Gyr. 
We see that the final distribution of GCs is slightly more centrally concentrated
in the second case. However, the value of $\mathcal{N}_{\Pi}$ in these two variations
is essentially the same as in the fiducial model.

To compute the evolution of the orbital radius of the GCs, we have used the
\citet{arc15} formula. To derive that formula, they carried out a careful 
modelling to include correctly the contribution to the drag from particles deep 
in the cuspy region of the galaxy. Still, we have explored the impact on the results 
if the dynamical friction timescale $\tau_{\rm \smalls DF}$ given in 
Equation (\ref{eq:tau_DF_Arca}) is doubled. Figure \ref{fig:twice_ArcaS}
shows the predicted radial distributions for such a case and assuming $\lambda=2$. 
We find that the number of galaxies with a central GC would be 
$8\substack{+2\\-1}$, which is close to the observed value. On the other hand, 
the predicted value of $\mathcal{N}_{\Pi}$ in this model is $4\substack{+3\\-1}$,
with a $4.5\times 10^{-4}$ probability that $\mathcal{N}_{\Pi}\geq 13$.
This reduction in the dynamical friction strength is not enough to reconcile models 
with the data. As expected, a reduction in dynamical friction implies a more compact
initial distribution of the GCs, i.e., a smaller value of $\lambda$. This does not result
in an enhancement of the correlation between
$\ell$ and $\mathcal{R}_{2\smalls\rm D}$; we find
a correlation coefficient $\hat{\rho}_{s}=-0.25\pm 0.12$ in this case.

In summary, we have adopted an NFW profile for the dark haloes of dSph galaxies
and find that the radial distribution of the GCs at distances 
$\mathcal{R}_{2\smalls \rm D}>0.5$ can be fitted reasonably well by assuming
a simple distribution for the starting GC distances. However, 
even adopting rather artificial starting positions for the GCs, 
such as truncated distributions,
it is difficult to account for the relatively high number of GCs inside 
$\mathcal{R}_{2\smalls \rm D}=0.5$, 
that are caught in their journey inwards. This is seen as a timing problem.
In Section \ref{sec:discussion}
we discuss the impact of some omitted processes, such as the stalling of 
the dynamical friction when the enclosed mass is comparable to the GC mass.
Before that, we present complementary views of the timing problem in the next
section.

\begin{figure*}
\hspace*{-0.5cm}\includegraphics[width=193mm, height=109mm]{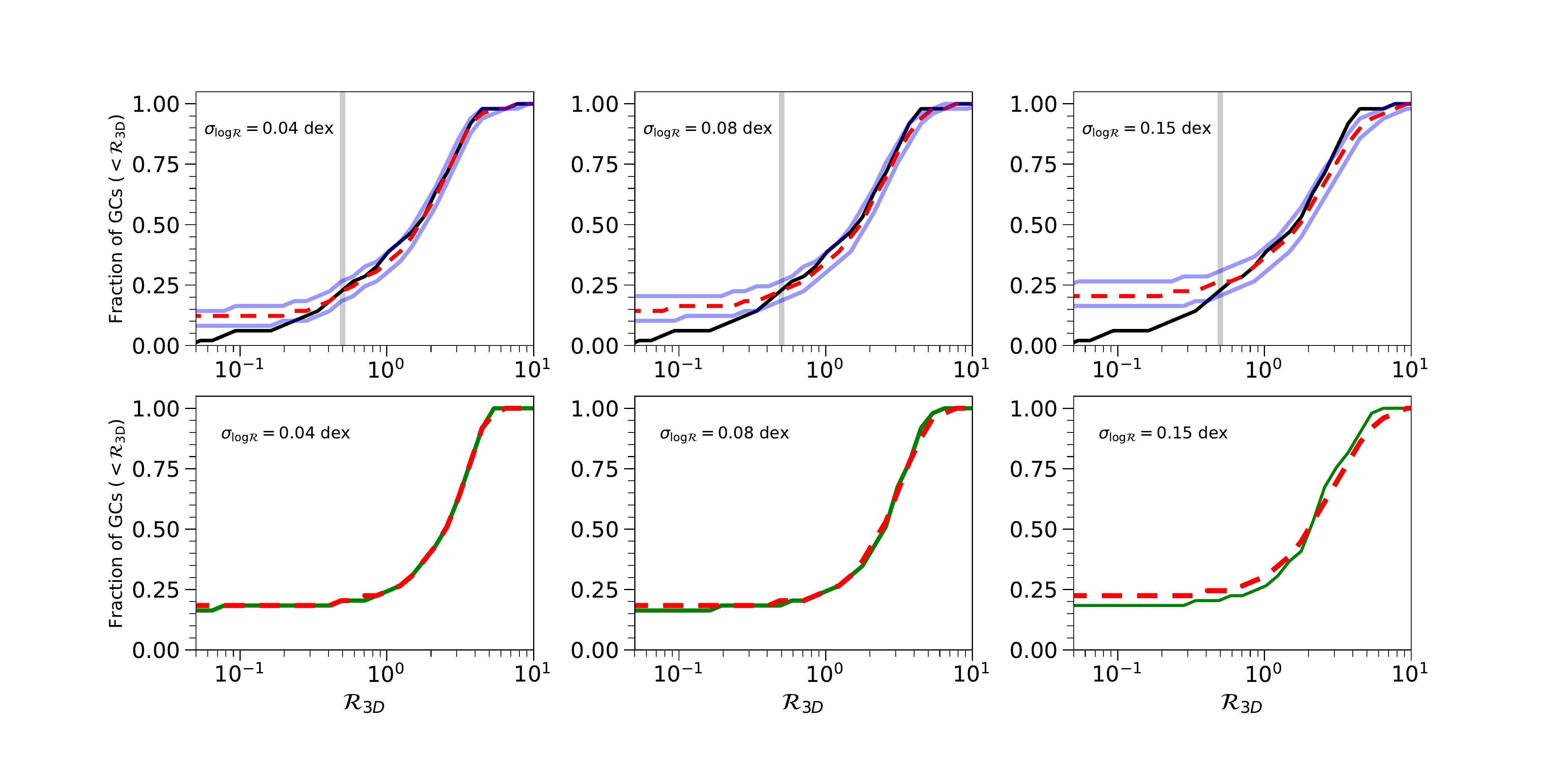}  
 \caption{Cumulative distributions of 3D distances of the non-central GCs at the 
present epoch. In the top panels, the black lines represent the 
present-day distribution from the data, the red dashed lines indicate the median of the
present-day distributions when the starting radii of the GCs are taken as
$\mathcal{R}_{3{\smalls \rm D},j}^{(i)}+\delta \mathcal{R}_{j}^{(i)}$, i.e. they are 
shifted from their original values in each trial, with 
$\sigma_{\smalls {\rm log}\mathcal{R}}=0.04$ dex (left panels), 
$0.08$ dex (central panels) and $0.15$ dex (right panels).
Blue lines indicate the $16$th and $84$th percentiles of the distributions.
As a guide for the eye, the vertical lines mark the radial distance 
$\mathcal{R}_{3\smalls \rm D}=0.5$.
In the bottom panels, we take a random set of $\mathcal{R}_{3{\smalls \rm D},j}^{(i)}$ 
using the PDF given in Equation (\ref{eq:NGC_Rinit}) with $\lambda=2.3$,
and compute the present-day cumulative distributions
without adding ``noise'' (green curves) and adding ``noise'' to the starting GC
distances (red dashed lines).
}
\raggedright
\label{fig:orbit_fine_tuning}
\end{figure*}

\begin{figure*}
\hspace*{-0.5cm}\includegraphics[width=195mm, height=54mm]{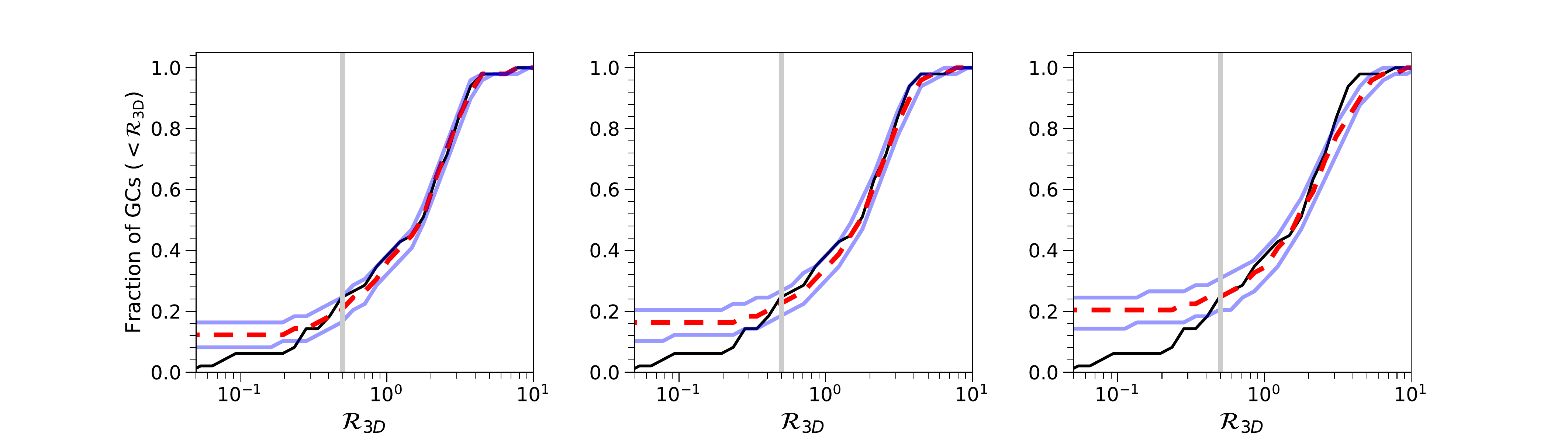}  
 \caption{Same as top panels in Figure \ref{fig:orbit_fine_tuning} but the values
of $R_{e}$, $(M_{\star}/L)_{\rm \smalls gal}, (M/L)_{\rm \smalls GC}, M_{200}, c$ and
$\tau_{\rm \smalls GC}$ have been fixed to their median values, i.e. they are not 
treated as random variables.
}
\raggedright
\label{fig:orbit_fine_tuning0}
\end{figure*}

\section{Past and future orbital radii of the GCs: 
Alternative views of the timing problem} 
\label{sec:past_future}
\subsection{Starting GC distances}
\label{sec:starting_dist}
In the previous section we have assumed the distribution of the starting orbital 
radii of the GCs and computed the present-day distribution. Then we have 
compared the
resultant radial configurations of the GCs with the observed ones. One benefit of
that approach is that projection effects are included 
in a natural way, because the Monte Carlo models contain information about 
both $3$D and projected radii of the GCs.

An alternative approach consists of computing the starting orbital radii from
the current projected distances of the GC.  To do so, we have to estimate
the present-day $3$D distance of each GC, $\mathcal{R}^{(p)}_{3{\smalls \rm D},j}$,
where the superscript $p$ indicates `present' and the index $j$ refers to the
$j^{\rm th}$ GC. In principle, it is possible to do that using deprojection techniques.
However, given the small number of GCs in our sample, we assume that
$\mathcal{R}^{(p)}_{3{\smalls \rm D},j}=\sqrt{3/2}R_{2{\rm \smalls D},j}^{(p)}$
\citep{mea20,sha21}.
For those GCs that are not at the centre
of the host galaxy (i.e. $\mathcal{R}^{(p)}_{3{\smalls \rm D},j}\neq 0$), 
we invert Equation (\ref{eq:final_radius_df}) to find the starting
distance $\mathcal{R}^{(i)}_{3{\smalls \rm D},j}$ using our Monte Carlo model.
For illustration, Figure \ref{fig:initial_radius_2GCs} shows the distributions 
of the present-day and starting distances for two GCs:  
KK 221 2-966 and KK 84 3-705. We note that the width of the distributions of
$\mathcal{R}^{(p)}_{3{\smalls \rm D},j}$ is exclusively due to the uncertainties in
the determinations of $R_{e}$.

Figure \ref{fig:initial_radius_allGCs} shows the PDF of the starting galactocentric
distances of all GCs excluding NSCs. The PDF presents local maxima, 
which are expected due to the small number of GCs in the sample. We see that the 
derived PDF is consistent with the analytical distribution used in
Section \ref{sec:modelA} (more specifically, Equation \ref{eq:NGC_Rinit} 
with $\lambda=2.5$).  The histogram in blue 
in Figure \ref{fig:initial_radius_allGCs} indicates that most of the non-central GCs
that are currently at projected distances $<0.5 R_{e}$ started their decay 
at 3D distances less than $3R_{e}$.

KK 221 2-966 is the GC that was excluded in the analysis in the previous section.
As expected, KK 221 2-966 requires to start its orbital decay at relatively large distances.
In our model, its starting 3D distance should be $5.25\pm 0.70$ galaxy effective
radius (see Figure \ref{fig:initial_radius_2GCs}), which corresponds to 
$3.25\pm 0.45$ kpc. This starting radius is similar to the estimated halo scale radius,
$r_{s}=3.1\substack{+0.90\\-0.65}$ kpc, in this galaxy.
For $\lambda=2.3-2.5$, the probability that
a GC starts at a distance $\geq 5 R_{e}$ is $(2.0-4.5)\%$.
Since we have $5$ GCs in our sample with $M_{V}<-9.5$, the probability that
at least one of these massive GCs started at a distance greater than $5R_{e}$
is $\sim (10-20)\%$. Thus, our model with $\lambda\simeq 2.3-2.5$ is
consistent with a scenario where KK 221 2-966 is a GC that was
caught still in-spiralling towards the galactic centre.

From the histogram of GC starting distances (Figure 
\ref{fig:initial_radius_allGCs}), we can derive the fraction of the GCs that
have initial 3D distances larger than $3R_{e}$.  Note, however,
that the present NSCs were not included in that histogram because their starting
orbital radius cannot be determined using Equation (\ref{eq:final_radius_df}). 
Since four of the NSCs are very luminous and 
dynamical friction timescale scales as $M_{\rm \smalls GC}^{-0.67}$, they could
have started their orbital decay at far distances from their galaxy centres.
If we suppose that the present NSCs started inside a sphere of radius $3R_{e}$,
we obtain that $36\%$ of the GCs started their orbital decay outside $3R_{e}$.
This value is a lower limit because some of the central GCs 
could have started at larger distances.
On the other hand, this fraction becomes $32\%$ if it is further assumed that 
dynamical friction stalls at $\sim 0.5R_{e}$ (see Section \ref{sec:discussion}).  
These estimated fractions seem quite large if we compare with the results 
of E-MOSAICS in \citet{sha21}, who report that about $20\%$ of the GCs in their Fornax
analogue sample have distances at birth, or at infall, larger than $3R_{e}$ 
(see their Figure 6),  where we have taken that $R_{e}\sim 1.3$ kpc in these Fornax-mass
dwarfs. However, it is worth noting that our distribution of starting orbital radii 
was obtained for those GCs that survived to tidal disruption, not for all the GCs 
as calculated by \citet{sha21}.
It should be also noted that $16\%$ of the GCs in our sample have present-day 3D distances larger than $3R_{e}$. As the gas density in the protogalaxies would be very low at 
those large distances, it is plausible that some of these GCs were not formed in situ 
but accreted, or they are not gravitationally bound to their respective dSph galaxies, 
or they are foreground/background GCs incorrectly associated to the dSph galaxies.

\subsection{The fine-tuning problem}

It is clear that given $\mathcal{R}^{(p)}_{3{\smalls \rm D},j}$, it is always
mathematically possible to find $\mathcal{R}^{(i)}_{3{\smalls \rm D},j}$, as did above. 
However, this fact does not mean that there is no a fine-tuning problem of 
the starting radii of GCs. To highlight the fine tuning problem,
we have carried out the following experiment. 
For each trial, we compute the starting orbital
radii of non-nuclear GCs, $\mathcal{R}^{(i)}_{3{\smalls\rm D},j}$, as we did in the
previous section.
Then, for each trial, we compute $\mathcal{R}^{(p)}_{3{\smalls \rm D},j}$
but starting with $\mathcal{R}^{(i)}_{3{\smalls \rm D},j}+\delta \mathcal{R}^{(i)}_{j}$,
where these initial orbital radii are lognormally sampled with $\sigma=0.04, 0.08$ 
and $0.15$ dex. Figure \ref{fig:orbit_fine_tuning} shows the median of the cumulative distribution 
$f_{\smalls \mathcal{N}}(\mathcal{R}_{3\smalls \rm D})$ when this exercise is
carried out for $4\times 10^{4}$ trials.
We see that these changes in the starting GC distance produce: (1) an
enhancement in the number of GCs that reach the centre of the galaxies,
and (2) a decrease in the number of GCs in the region 
$0.05\leq \mathcal{R}_{3\smalls \rm D}\leq 0.5$. For instance, for $\sigma=0.08$ dex,
which implies fractional changes in the starting radii
of $\pm 20\%$, the number of GCs that sink to
the centre of galaxies increases by $9$, whereas the number of GCs inside
$\mathcal{R}_{3\smalls \rm D}=0.5$ (excluding the nuclei)
decreases from $11$ to $3$. On the contrary,
if this exercise is done not for the observed $R_{2{\rm \smalls D},j}^{(p)}$ but for
a random set, we find that $f_{\smalls \mathcal{N}}(\mathcal{R}_{3\smalls \rm D})$
is not sensitive to changes in the orbital configuration provided that 
$\sigma\lesssim 0.15$ (see bottom row in Figure \ref{fig:orbit_fine_tuning}).

In order to isolate the effect of changes in the initial starting position of the GCs,
we compute the distributions adding noise to $\mathcal{R}^{(p)}_{3{\smalls \rm D},j}$,
but fixing the values of 
$R_{e}$, $(M_{\star}/L)_{\rm \smalls gal}$, $(M/L)_{\rm \smalls GC}$, $M_{200}$, $c$ 
and $\tau_{\rm \smalls GC}$ to their median values. The results are displayed in
Figure \ref{fig:orbit_fine_tuning0}. 
We see that the scatter of
$f_{\smalls \mathcal{N}}$ (blue lines) is essentially the same as it is when
those variables are not fixed. This indicates that the noise introduced in the 
starting GC radii primarily determines the scatter of $f_{\smalls \mathcal{N}}$.

To summarize, the inverse calculation of the GC orbits allow us to derive their 
starting orbital radii, from their current positions. In fact, we obtain a smooth
distribution of starting positions.
However, if we sample this distribution and calculate the expected present-day
$f_{\smalls \mathcal{N}}$, we do not recover the observed distribution but 
obtain a shallower profile at $\mathcal{R}_{3\smalls \rm D}\leq 0.5$, with a 
central flat region. 
The width of this central flat region depends on how much 
the initial radii of the GCs differ from the exact initial starting position required.

\subsection{Future distribution of GCs}
\label{app:future3Gyr}

The possible connection between GCs and  NSCs has been a topic of active research
\citep{san19,hoy21,car21}. It was found that the nucleation fraction (i.e. the fraction
of galaxies with a NSC) and the GC occupation fraction (i.e. the fraction of galaxies 
with one or more GCs) for dwarf galaxies vary similarly with stellar mass in Virgo 
and Coma clusters but also in the Local Volume. This result has been interpreted
as suggestive that the formation of NSCs in these dwarf galaxies are dominated   
by the in-spiral and merger of GCs (although
NSCs could also grow through in-situ star formation after forming).
For galaxies with stellar mass $\sim 10^{7}M_{\odot}$, the nucleation fraction is 
$\simeq 0.3-0.4$ in Virgo and Coma cluster, and $\sim 0.2$ in the Local Volume
\citep{san19,hoy21,car21}.

\begin{figure}
\hspace*{0.0cm}\includegraphics[width=83mm, height=131mm]{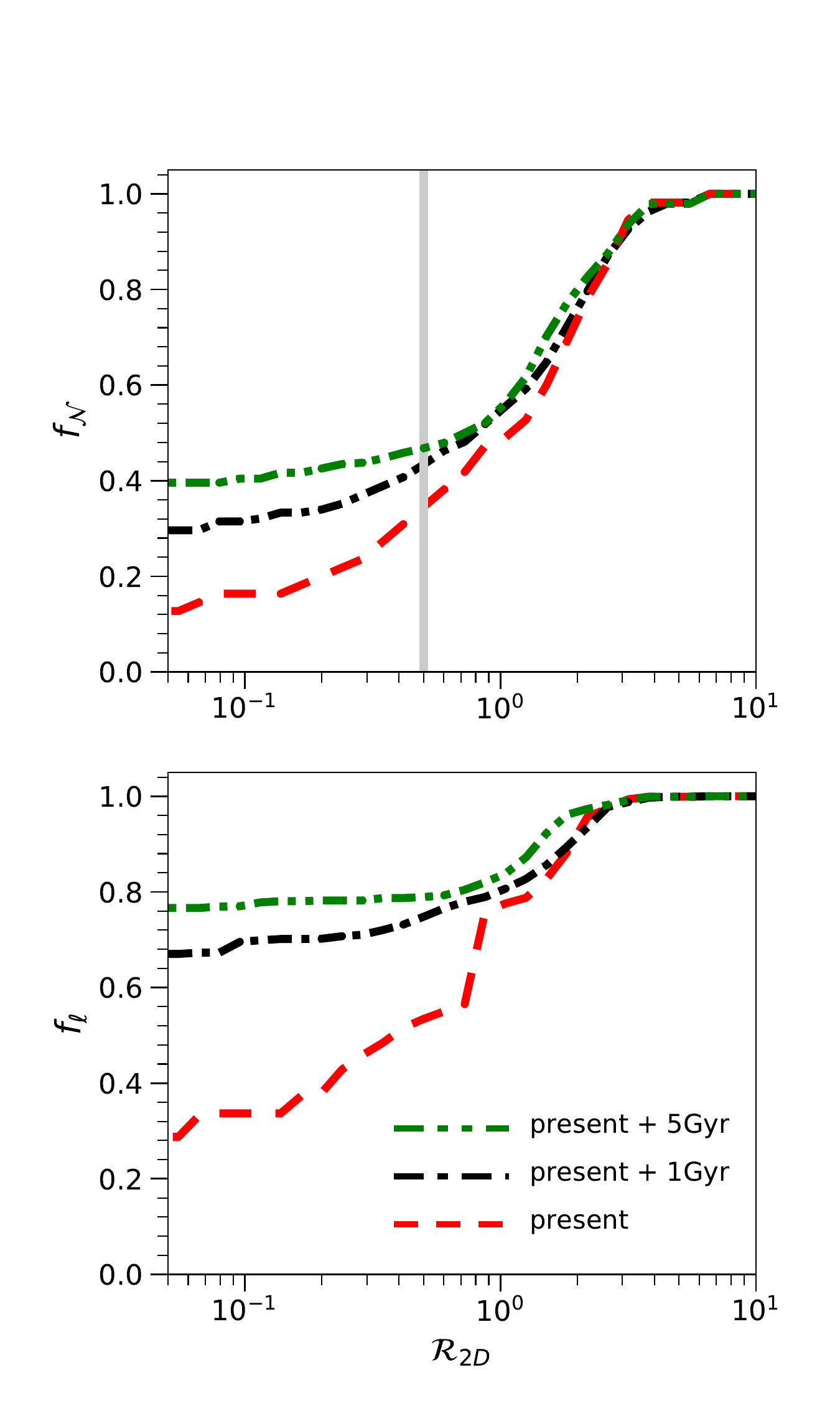}  
 \caption{$f_{\smalls \mathcal{N}}$ (top panel), and  
$f_{\ell}$ (bottom panel), as a function of projected radius, at three different times 
into the future, assuming that all GCs survive against tidal disruption.
The full sample of GCs was used. As a guide for the eye, the vertical line in
the upper panel marks the radial distance $\mathcal{R}_{2\smalls \rm D}=0.5$.
}
\raggedright
\label{fig:future}
\end{figure}

\begin{figure}
\hspace{0.2cm}
\includegraphics[width=80mm, height=61mm]{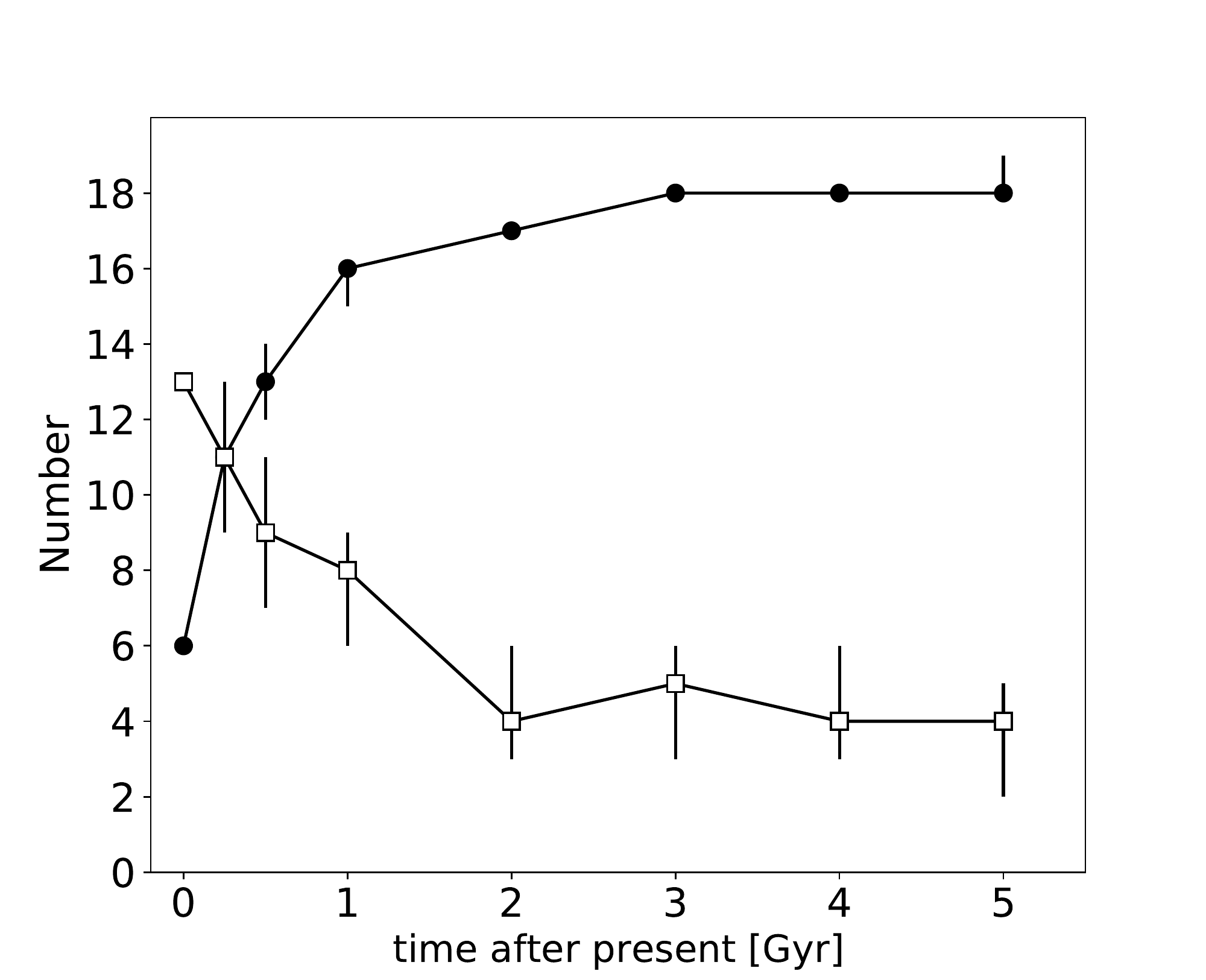}
 \caption{Number of galaxies with an NSC (filled circles) and $N_{\Pi}$
(empty squares) over the next $5$ Gyr, assuming that all GCs survive against tidal disruption.  We have used the full sample of GCs.
}
\centering
 \label{fig:N_time}
 \end{figure}

In this context, the nucleation fraction should increase with time as GCs sink to the galaxy
centre. From the observed current values of $M_{\smalls V,\rm GC}$ 
and $\mathcal{R}_{2\smalls \rm D}$ of the GCs in our sample, 
using our Monte Carlo model described in Section \ref{sec:MC_model},
we can predict the radial distribution of GCs at later times. 
Figure \ref{fig:future} shows $f_{\smalls \mathcal{N}}$ and $f_{\ell}$ at present
time, at $1$ Gyr, and at $5$ Gyr from now.
It was assumed again that the current 3D distance of the GCs are $\times \sqrt{3/2}$ 
their projected distance, and that all the GCs survive against tidal disruption
for the next $5$ Gyr. The redistribution of the GCs inside one effective radius of 
the galaxies is apparent. Both profiles, $f_{\small \mathcal{N}}$ and $f_{\ell}$, flatten 
in the inner region because there dynamical friction is very efficient.
The number of galaxies hosting a NSC will change from $6$ ($23\%$ of the
dSph galaxies under study) at the present epoch to $16$ ($60\%$) in the next $1$ Gyr 
(see Figure \ref{fig:N_time}). On the other hand,
Figure \ref{fig:N_time} also shows that 
the number of GCs inside $\mathcal{R}_{2\smalls \rm D}=0.5$ decreases from
$\mathcal{N}_{\Pi}=13$ to $4$ in the next $2$ Gyr.
Since the present distribution of GCs evolves in a short 
timescale compared with the lifetime of the GCs, it would imply that we are living in a 
special time of their evolution. \citet{col12} refers to this problem
as the immediate timing problem. This timing problem is a different view of the
problem of fine-tuning of the starting distances of the GCs.

\section{Discussion}
\label{sec:discussion}
\subsection{Eccentric orbits, tidal disruption of GCs, and tidal stripping of dark haloes}
The results of the previous sections rely on various assumptions and simplifications.
For all clusters, we have adopted circular orbits. Some of the GCs that are now
located at $0<\mathcal{R}_{2\smalls \rm D}<0.5$ may have avoided
strong orbital decay if they are on elongated orbits and they are now
passing close to pericentre. Given that GCs in eccentric orbits spend a considerable
fraction of the orbital time far from pericentre, 
the inclusion of eccentric orbits for outer GCs 
can alleviate slightly the tension between predictions and observations. 
In addition, if inner GCs are on elongated orbits, the problem is exacerbated because
eccentric orbits have a shorter $\tau_{\smalls \rm DF}$ according
to \citet{arc15} formula (see Section \ref{sec:MC_model}).

We must stress that our Monte Carlo simulations do not include disruption 
of GCs by tidal forces. Because of the tidal dissolution of GCs, the current population 
of GCs is a fraction of the initial population. 
Given that the strength of the tidal forces on GCs in cuspy haloes 
increases as they sink towards the galactic centre, we would expect 
a depletion of GCs in the inner galaxy \citep[e.g.,][]{sha21}. 
Therefore, it seems challenging that the inclusion of tidal disruption of GCs can 
solve the offset in the value of $\mathcal{N}_{\Pi}$.
Nevertheless, it could be interesting to
start with a larger population of GCs and derive the final spatial distribution
of GCs when tidal disruption is included.

\begin{figure*}
\vspace*{-0.4cm}
\includegraphics[width=130mm, height=203mm]{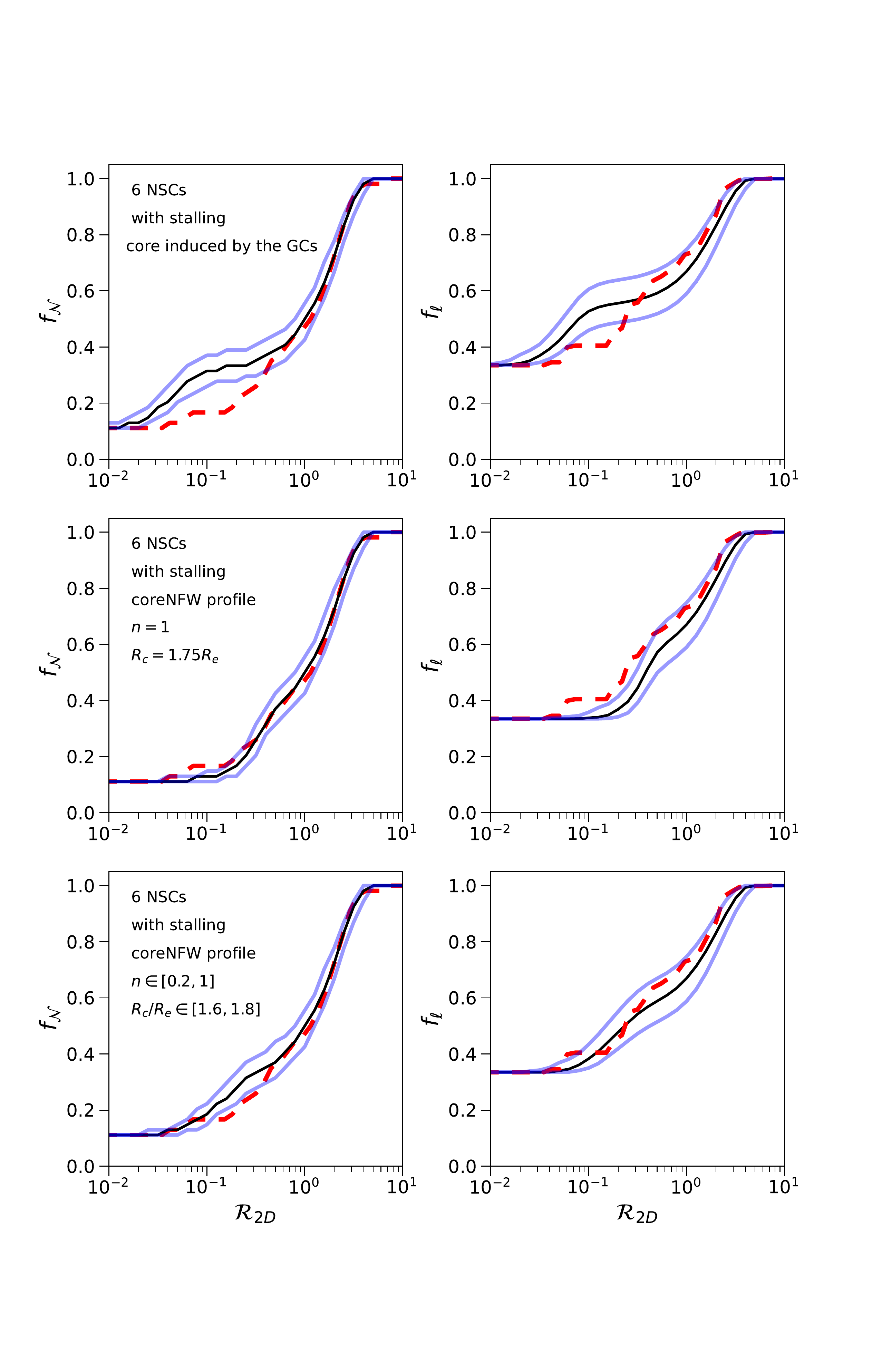}  
\vspace*{-1.3cm}
 \caption{Comparison between the predicted distributions
$f_{\smalls \mathcal{N}}$ and $f_{\smalls \ell}$ in cored dark matter haloes and 
those determined by observations.
We assume that the inspiralling of GCs stalls at the radius $R_{\rm stall}$, which
satisfies $R_{t}(R_{\rm stall}) = R_{\rm stall}$, being $R_{t}$ the GC tidal radius. 
The tidal radius is computed using either the NFW profile or
the coreNFW profile with parameters $n$ and $R_{c}$ as indicated in the left insets.
The six central GCs are treated as `pure' NSCs,
i.e.~it is assumed that they were born at the centre of the host galaxy.   
The distribution of starting radial distances of non-central GCs is given by Equation (\ref{eq:NGC_Rinit}) as in our canonical model. Line colours and style codes
are the same as those used in Figures \ref{fig:without_1GCs}, \ref{fig:initially_segregated},
\ref{fig:6NSC_no_stalling},
\ref{fig:truncated_Gaussian}, and \ref{fig:twice_ArcaS}.
}
\centering
\label{fig:6NSC}
\end{figure*}

We have also ignored the effect of the tidal heating of 
the host galaxies by companion galaxies. 
Tidal stripping and shocking by neighbouring galaxies can lower the central dark 
matter density of the haloes of dSph galaxies \citep[e.g.,][]{bat15,gen20}. 
The values of the tidal index $\Theta$ in Table \ref{table:dSph_properties}
indicate that tidal effects can be important
for some dSph galaxies in our sample. In particular, And XXV has peculiar properties,
which were interpreted as the result of tides \citep{col13}.
The strength of tidal fields depends on the pericentre of the 
dSph's orbits around the
host galaxy, which is unknown for most of the galaxies in our sample.
In the case of Fornax, tidal stirring has been 
considered by other authors and their effects
seem to not alter significantly the orbital decay of its GCs 
\citep{oh00,arc16,bor21}. Nonetheless, these issues warrant more attention
in future work.

\subsection{Core stalling as a solution to the timing problem}

The timing problem (or fine-tuning problem) could be interpreted as indications 
of the breakdown of the NFW profile at small radii. 
Indeed, GCs themshelves can tidally dirupt the dark halo cusp,
transforming it into a core, when the sinking GCs are close to the galaxy 
centre \citep{goe10}.
This transformation occurs when the orbital radius of the GC is comparable to its
tidal radius. When a core is formed, dynamical friction is suppressed and GCs 
stall. \citet{goe10} were able to empirically derive the 
core-stalling radius in terms
of the distance at which the mass of the satellite equals the enclosed mass of the host
galaxy. However, \citet{pet16} find a more simple prescription for core-stalling
in terms of the tidal radius of the GC; core-stalling happens at a radius $R_{\rm stall}$
given by 
\begin{equation}
R_{t}(R_{\rm stall})=R_{\rm stall},  
\label{eq:stalling_condition}
\end{equation}
where
\begin{equation}
R_{t}^{3}= \frac{G M_{\rm \smalls GC}}{\Omega^{2}-d^{2}\Phi/dr^{2}},
\end{equation} 
with $\Phi$ the gravitational potential of the host galaxy and $\Omega$ 
the angular velocity of a test particle in
circular orbit, both evaluated at the instantaneous orbital radius of the GC.

We have explored how $f_{\smalls \mathcal N}$ and $f_{\ell}$ are modified
in this scenario as follows. We assume that
the host galaxies follow an NFW profile before cusp-core transformation, with
the halo parameters as described in Section \ref{sec:MC_model}. Then, 
for each host galaxy, we know $\Omega (r)$ and $d^{2}\Phi/dr^{2}$.
For each non-central GC and its host galaxy, we determine
$R_{\rm stall}$ by solving Equation (\ref{eq:stalling_condition}).
Finally, when computing the model distributions, we switch off dynamical friction at 
$R_{\rm stall}$. For the specific six GCs that are observed at the centre of the host galaxy
(see Section \ref{sec:variations}), we assume that they were formed at the galaxy centres,
i.e. they are pure NSCs ($\mathcal{R}_{3\rm \smalls D}=0$ at any time).
Otherwise, this scenario would predict that no GC could lie in the centres
of galaxies, as they would always halt their migration at $R_{\rm stall}$.

The top panels in Figure \ref{fig:6NSC} show the results. By construction, the values of 
$f_{\smalls \mathcal{N}}(0)$ and $f_{\ell}(0)$ match the observed values because of the
assumption that NSCs should have been formed at $r=0$. The inclusion
of core-stalling leads to steeper profiles inside $0.1 R_{e}$, which reflects the fact
that typically $R_{\rm stall}\lesssim 0.1R_{e}$ in our sample. However, the observed
profiles are almost flat in that region. In fact, the cores formed by the GCs in their own
are too small to satisfactorily account for the observed mean abundance and luminosity
of GCs between $0.1R_{e}$ and $0.7 R_{e}$. Almost identical diagrams are
obtained if instead of the condition $R_{t}(R_{\rm stall})=R_{\rm stall}$, 
we assume that stalling occurs at a radius where the enclosed mass is comparable to 
the GC mass (not shown).

It has been suggested that gravitational potential fluctuations, created by gas outflows 
driven by bursty star formation and episodes of gas inflows, 
could be a viable mechanism to transform cusps into cores 
\citep[e.g.,][]{pon12,rea16,laz20}. 
Core formation can also be driven by angular momentum transfer between
dark matter particles and cold gas clumps \citep{elz01,nip15}, or through
impulsive heating from minor mergers \citep{ork21}. Other scenarios
for core formation are based on modifications to the dark matter physics
\citep[e.g.,][]{dav01,san03,sch14}.

Figure \ref{fig:6NSC} shows the expected radial distributions 
using the ``coreNFW'' profile from \citet{rea16} for all the galaxies in the sample. 
The coreNFW profile is a modified NFW profile which was found to describe the 
simulated dark haloes in dwarf galaxies after including stellar feedback. 
The model is parameterised by the size of the dark halo core $R_{c}$ and 
the power-law slope of the core $0\leq n\leq 1$, where $n=0$ corresponds to
the cuspy NFW profile and $n=1$ produces a flat dark matter core. As before, the 
stalling radius is computed by solving the Equation (\ref{eq:stalling_condition}).

We find good agreement between the predicted and the observed
distributions when we set $n$ to $1$ and $R_{c}$ to $1.75R_{e}$, as suggested 
by \citet{rea16}. 
We also show the results in a case where $n$ is uniformly
sampled between $0.2$ and $1$, and $R_{c}$ between $1.6R_{e}$ and $1.8R_{e}$. 
The agreement between the observed and the predicted profiles is reasonable.
Therefore, the hypothesis that the dark haloes
of dSph galaxies have cores of size $\sim R_{e}$ is a natural solution for
the timing problem; both the low fraction of central GCs and the high
fraction of off-centre GCs inside $0.5R_{e}$ could be a consequence of the 
reduction of dynamical friction inside the cores. 
Note, however, that dark matter cores of size comparable to $\sim R_{e}$ can form 
only if star formation proceeds for several Gyr \citep{rea16}, and star formation 
is bursty enough to prevent that subsequent gas accretion reforms a cusp \citep{ben19}.

\section{Conclusions}
\label{sec:conclusions}
In this paper, we have collected from the literature a sample of GCs hosted by 
low-luminosity spheroidal galaxies, in order to investigate the role of dynamical 
friction in the spatial distribution of the GCs in these galaxies.
Firstly, we have searched for any statistical evidence of mass
segregation in the stacked distribution of GCs. We have found a moderate correlation 
between specific luminosity ($\ell$) and the projected distance of the GCs to the centre of 
the host galaxy in units of the core radius ($\mathcal{R}_{2\smalls \rm D}$), with a high 
significance level.

We have performed simple Monte Carlo simulations to predict the radial distribution
of the number and luminosity of the GCs, assuming that their orbital evolution is driven 
by dynamical friction with the dark matter.
If all dSph galaxies in our sample have an NFW dark halo, 
the predicted number of non-nuclear GCs inside one half effective radius 
($\mathcal{N}_{\Pi}$) is considerably smaller than observed 
(see Figure \ref{fig:central_periphery}). In the models,
$\mathcal{N}_{\Pi}$ is low because GCs spend little time in that inner region given
that the timescale to sink is very short. Therefore, the timing problem of the orbital
decay of GCs is not exclusive of the Fornax dSph galaxy.

The timing problem is equivalent to invoke a fine-tuning in the starting radii
of the GCs. In fact, it is always possible to find the starting distance of the GCs if
we know the final orbital radii and the halo parameters (see
Figure \ref{fig:initial_radius_allGCs}), but relatively small changes
in the starting distances lead to very different final orbital radii. 

A third view of the timing problem is the ``immediate timing problem'' \citep{col12}.
Using our probabilistic approach and the observed present-day GC projected distances, 
we have predicted the future radial distribution of GCs, assuming that all GCs survive
against tidal dissolution (see Figure \ref{fig:future}). 
While at the present-day, about
$\sim 23\%$ (6/26) of the galaxies in our sample has a central GC, this fraction will
change to becoming $\sim 60\%$ (16/26) in the next $1$ Gyr. In addition, the value of 
$\mathcal{N}_{\Pi}$ will change from $13$ at present-day to $4$
at $2$ Gyr into the future.

The timing problem is alleviated if, instead of an NFW profile, one considers weakly
cuspy haloes or cored haloes. In particular, the present-day GC distribution
can be accounted for if dark haloes have cores of size $\sim R_{e}$ 
(see Figure \ref{fig:6NSC}). Several mechanisms have been
suggested that could transform cusps into cores. Thus, the possibility that the NFW
profile breaks down at small radii in dSph galaxies seems plausible. It is less
clear whether gas outflows/inflows can develop cores of that size.

Studies including the internal dynamics and the survival of GCs 
may provide additional constraints on the dark matter profile \citep[e.g.,][]{amo17,
con18,ork19,lea20}. 
Moreover, new data of the distribution of GC candidates around early-type dwarf galaxies
\citep[e.g.,][]{car21} could significantly increase the statistics for studying 
the inspiralling of GCs and the role of dynamical friction. 
This investigation is planned for future work.

\section*{Acknowledgements}
We are indebted to the anonymous referee for a deep scrutiny of the paper with
very helpful comments and suggestions that improved our work.
This work was partially supported by PAPIIT project IN111118.

\section*{Data Availability}
No new data were generated or analysed in support of this research.


\bsp	
\label{lastpage}
\end{document}